\documentclass[twocolumn]{aastex62}

\usepackage{amsmath,amssymb,gensymb,times,graphics,morefloats,color}

\newcommand{\vsini}{$v\sin i$~}
\newcommand{\lande}{Land\'e~}
\newcommand{\ti}{\ion{Ti}{1}~}
\newcommand{\ca}{\ion{Ca}{1}~}

\newcommand{\nirspectargets}{30~}
\newcommand{\carmenestargets}{44~}
\newcommand{\totaltargets}{70~}

\received{September 24, 2019}
\accepted{November 27, 2019}
\submitjournal{AJ}

\shorttitle{Absorption vs. Rossby Number in M dwarfs}

\shortauthors{Muirhead et al.}

\begin{document}

\title{Magnetic Inflation and Stellar Mass. V. Intensification and saturation of M dwarf  absorption lines with Rossby number}

\correspondingauthor{Philip S. Muirhead}
\email{philipm@bu.edu}

\author[0000-0002-0638-8822]{Philip S. Muirhead}
\affil{Department of Astronomy, Institute for Astrophysical Research, Boston University, 725 Commonwealth Avenue, Boston, MA 02215, USA}

\author[0000-0002-0385-2183]{Mark J. Veyette}
\affil{Department of Astronomy, Institute for Astrophysical Research, Boston University, 725 Commonwealth Avenue, Boston, MA 02215, USA}

\author[0000-0003-4150-841X]{Elisabeth R. Newton}
\affil{Department of Physics \& Astronomy, Dartmouth College, Hanover, NH 03755,USA}


\author[0000-0002-9807-5435]{Christopher A. Theissen}
\altaffiliation{NASA Hubble Fellowship Program Sagan Fellow}
\affil{Center for Astrophysics and Space Sciences, University of California, San Diego, 9500 Gilman Dr., La Jolla, CA 92093, USA}

\author[0000-0003-3654-1602]{Andrew W. Mann}
\affil{Department of Physics and Astronomy, University of North Carolina at Chapel Hill, Chapel Hill, NC 27599, USA}

\begin{abstract}

In young sun-like stars and field M dwarf stars, chromospheric and coronal magnetic activity indicators such as H$\alpha$, X-ray and radio emission are known to saturate with low Rossby number ($Ro \lesssim 0.1$), defined as the ratio of rotation period to convective turnover time.  The mechanism for the saturation is unclear.  In this paper, we use photospheric \ti and \ca absorption lines in $Y$ band to investigate magnetic field strength in M dwarfs for Rossby numbers between 0.01 and 1.0.  The equivalent widths of the lines are magnetically enhanced by photospheric spots, a global field or a combination of the two.  The equivalent widths behave qualitatively similar to the chromospheric and coronal indicators: we see increasing equivalent widths (increasing absorption) with decreasing $Ro$ and saturation of the equivalent widths for $Ro \lesssim 0.1$. The majority of M dwarfs in this study are fully convective.  The results add to mounting evidence that the magnetic saturation mechanism occurs at or beneath the stellar photosphere.


\end{abstract}


\section{Introduction\label{sec:intro}}

Magnetic activity of low-mass stars, specifically M dwarf stars, is of significant interest to multiple areas of astrophysics.  Magnetized winds are expected to be the dominant rotational breaking mechanism in main-sequence M dwarfs stars \citep[e.g.][]{Bouvier2014} as well as the dominant orbital breaking mechanism in post-common envelope binary stars \citep[e.g.][]{Muirhead2013} and cataclysmic variable stars \citep[e.g.][]{Skinner2015}, both of which often contain M dwarf stars.  Magnetic heating of M dwarf chromospheres and coronae results in  high-energy radiation \citep[e.g.][]{Stelzer2016, France2016}  and likely also coronal mass ejections \citep[e.g.][]{Kay2016}, all of which have implications for the atmospheres and surfaces of orbiting extrasolar planets.

Magnetic activity, broadly defined, is commonly probed using a handful of observational signatures, mostly through its effects on the stellar upper atmosphere.  The equivalent widths of H$\alpha$ or \ion{Ca}{2} emission lines are common tracers for magnetic activity, originating in the chromospheres of stars, as they do in the Sun \citep[e.g.][]{Noyes1984,Soderblom1993}.  Another signature of magnetic activity is X-ray emission, originating in the hot corona \citep[e.g.][]{Pizzolato2003}, as well as radio emission from electrons accelerated in the stellar magnetic field \citep[e.g.][]{Stewart1988,Berger2006}.  Though, a random measurement of a chromospheric emission line or X-ray emission may capture an active flare and not be representative of a star's quiescent state \citep[e.g.][]{Paulson2006}.  An interesting feature of these magnetic activity signatures is the saturation with stellar Rossby number $Ro$, defined as the ratio of rotation period to the convective turnover time.  For $Ro \gtrsim 0.1$, stars show a log-linear relationship between the strength of magnetic indicators, normalized to stellar luminosity, and Rossby number; however, for $Ro \lesssim 0.1$, the relationship is flat.  As $Ro$ probes magnetic field generation in the rotating and convecting stellar atmosphere, the saturation mechanism takes place somewhere in the rotating convection zone. The saturation is observed in sun-like stars \citep[e.g.][]{Pallavicini1981, Wilson1966} as well as M dwarfs stars on either side of the fully convective boundary \citep[][]{Newton2017, Wright2018}.

The nature of the saturation mechanism in unclear.  Proposed scenarios include centrifugal stripping of the corona \citep[][]{Jardine1999}; however, this scenario does not explain saturation in the chromosphere and has since become a preferred explanation for super-saturation seen for $Ro \lesssim 0.01$ in X-rays.  Other proposed mechanisms include reaching a maximum filling factor of active regions in the photosphere or saturation of the dynamo mechanism itself \citep[][]{Vilhu1984}.

If the saturation mechanism were a process confined to the stellar chromospere or corona, we would not expect {\it photospheric} magnetic activity indicators to saturate as well.  However, \citet[][]{Reiners2009} used Zeeman splitting in photospheric FeH lines of M dwarfs to show that the average unsigned magnetic field in the photosphere saturates as well.  This suggest that the saturation mechanism lies at or beneath the stellar photosphere, and not in the chromosphere or corona.  The investigation used rotational broadening of absorption lines instead of photometric rotation periods to estimate Rossby number and required an interpolation method to convert the FeH line profiles to an estimate of the unsigned magnetic field strength.  Recently, \citet{Shulyak2019} measured unsigned average magnetic field strength using \ti lines in $z$ band using data from the CARMENES exoplanet survey \citep[][]{Quirrenbach2016}.  They used photometric rotation periods and compared the magnetic field strengths derived using FeH lines to the strength derived from $z$-band \ti lines.  They found that at high \vsini it is difficult to determine the magnetic field strength from FeH lines alone due to degeneracies in the derived parameters.  However, they did not investigate the magnetic field strength as a function of Rossby number, and nearly all of the targets have $Ro<0.1$, making it difficult to establish whether magnetic fields in the photosphere saturate across the $Ro \sim 0.1$ saturation point.

In this paper, we report on the magnetic sensitivity of \ti and \ca absorption lines in $Y$ band, from 10000 to 11000 Angstroms, versus Rossby number.  The $Y$-band spectral region is entirely free of telluric absorption features, which typically plague infrared spectroscopy.  We observed M dwarf stars at high spectral-resolution in $Y$ band using the NIRSPEC spectrograph on the Keck II Telescope, and we supplement our data with publicly available spectra from the CARMENES survey.  The targets have a range of rotation periods, masses and Rossby numbers.  We find that the equivalent widths of these lines saturate in {\it absorption} for $Ro<0.1$, similar to what is seen for X-rays and H$\alpha$ in emission.  The lines are magnetically enhanced (made {\it deeper}) due to the effect of Zeeman splitting on the curve of growth \citep[see, for example, ][]{Basri1992}. Our results provide further evidence that the fundamental magnetic saturation mechanism lies at or beneath the stellar photosphere.

\section{Data}\label{sec:data}

We collected spectra using one night of observations with the with the Near-Infrared Echelle Spectrograph (NIRSPEC) on the 10-meter Keck II Telescope located on the summit of Mauna Kea in Hawaii \citep[][]{McLean1998}, and combine them with publicly available data from the CARMENES M-dwarf survey \citep{Reiners2018}.  NIRSPEC is a cross-dispersed echelle spectrograph covering 1.0 to 5.0 $\mu$m in multiple settings, with a maximum resolving power ($\lambda/\Delta\lambda$) of 25000.  Wavelength regions within 1.0 to 5.0 $\mu$m are selected by the combination of a filter wheel and tilting/rotating motors mechanically connected to the echelle and cross-dispersing grating.  In 2018, NIRSPEC was upgraded to improve slit-viewing, increase overall sensitivity, and increase the simultaneous wavelength coverage in a single setting \citep[][]{Martin2018}.  Most relevant to this work, the original 1024x1024 pixel InSb detector (20-$\mu$m pixels) was replaced with a larger and more sensitive 2048x2048 pixel HgCdTe detector (15-$\mu$m pixels).  For this survey, we used the upgraded version of NIRSPEC on the night of UT 18 December 2019, the very first night of facility science operations of the upgraded instrument.  The upgrade required some alternations to existing data reduction methods, which are described in Section \ref{subsec:reduction}.  

\begin{figure}
\includegraphics[width=0.49\textwidth]{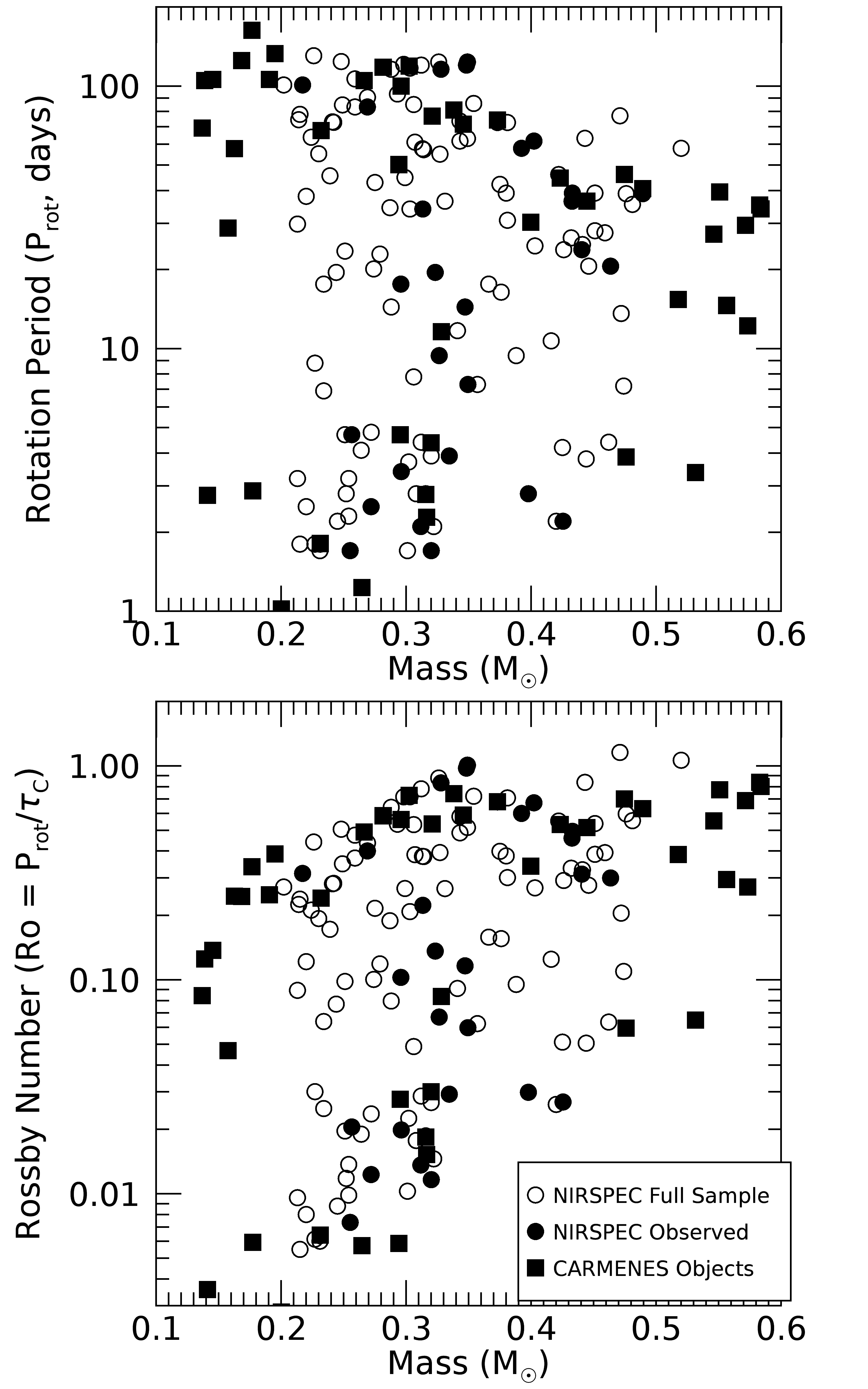}
\caption{Mass versus rotation period and Rossby Number for the objects in the full sample and those observed.  Masses were calculated using the mass-luminosity relations of \citet{Mann2019}, rotation periods are from \citet{Newton2017}, and convective turnover time was calculated using relations from \citet{Wright2011}.\label{fig:sample}} 
\end{figure}

\subsection{NIRSPEC Target Selection}

Targets for the NIRSPEC survey were chosen from \cite{Newton2017}, who combined photometric rotation periods of M dwarf targets in the MEarth transiting exoplanet survey \citep[][]{Charbonneau2009} with photometric rotation periods from the literature.  We required targets to have a $J$ magnitude of less than 11 and visibility from Mauna Kea.   We then combined the measured rotational period with stellar radius, using the $M_{Ks}$-to-radius relationship from \citet{Mann2015}, to determine the maximum possible \vsini.  We combined trigonometric parallax measurements from ESA's \textit{Gaia} Mission \cite[][]{Gaia2018} with $Ks$-band photometry from the TwoMicron All Sky Survey \citep[2MASS,][]{Skrutskie2006} to determine $M_{Ks}$ for each target. We limited our targets to those with a maximum possible \vsini of 12 km s$^{-1}$, corresponding to the instrumental resolving power of NIRSPEC.  We did this so that the rotational broadening would not significantly affect the resulting spectra. This resulted in 102 targets for the NIRSPEC sample.  We then estimated their masses using the $M_{Ks}$-to-mass relationship from \citet{Mann2019}.

The sample of M dwarfs from which we draw our targets for the NIRSPEC observations has been subject to substantial investigation by the MEarth team to identify known binaries. In that sample, data from the literature was used to identify eclipsing, visual, or spectroscopic binaries. Eclipsing binaries identified in the MEarth photometry \citep[e.g.][]{Dittmann2017,  Winters2018} and in spectroscopic data identified through follow-up of MEarth targets \citep{Newton2017} were noted. Stars overluminous for their color or spectroscopic properties based on relations from \citet{Newton2015} and \citet{Dittmann2016} were also noted as potential binaries.  We selected only those stars not tagged as binaries or potential binaries. We note that the stellar binary fraction of M dwarfs \citep[27\%][]{Winters2019} is significantly lower than for higher mass stars.

\subsection{NIRSPEC Observations}

Of the 102 identified for NIRSPEC observations, we observed \nirspectargets on UT 2018 December 18.  Conditions were photometric with an average seeing of 1\farcs{6}.  We used the NIRSPEC-1 mode and the 0\farcs{432} x 12\farcs{00} slit, which covers $Y$ band, roughly 9800 to 11000 Angstroms, at a resolving power of $R=\lambda/\Delta\lambda=25000$.  We acquired spectral dark images, flat-field images and wavelength-calibrating arc-lamp images following the standard NIRSPEC observing procedure at the beginning of the night.  We achieved a signal-to-noise ratio of between 60 and 100 across $Y$ band for each target.  

In addition to the \nirspectargets M dwarf targets, we also acquired spectra of a rapidly rotating A0 star, HD 43607, to provide a featureless spectrum for general calibration purposes.  We observed all targets following an ABBA dither pattern on the slit, and read out the detector using multiple-correlated non-destructive double sampling \citep[i.e. Fowler sampling, ][]{Fowler1990}.  Subtracting the AB (and BA) pairs removes detector fixed patterns and slowly varying background emission.

\subsection{NIRSPEC Reduction\label{subsec:reduction}}

We reduced the data using a combination of the NIRSPEC Data Reduction Pipeline (\texttt{NSDRP}) software package \citep{Tran2016}, updated for the upgraded version of NIRSPEC, and custom routines.  The \texttt{NSDRP} software package accepts spectral flat fields, arc-lamp images and an AB pair of science observations, and returns a series of reduced data products.  The software performs dark subtraction on the flat-field images, and divides the subtracted AB pair by the resulting flat-field image.  Following the flat fielding, the software operates on each spectral order independently, rectifying the echellegram to remove curvature of the order and the slit tilt, producing a mean slit profile for each order and summing pixels across the regions of the slit profile that contain the target (one region for the A image, one region for the B image) to produce spectra, then combines the A and B spectra into a single spectrum.  The software then processes the arc-lamp images to provide a wavelength solution for each order.  The program returns a two-dimensional rectified image, profile, spectrum and uncertainty for each order.

To improve signal-to-noise and bad-pixel rejection, we did not use the spectra returned by \texttt{NSDRP} and instead performed ``optimal extraction'' on the two-dimensional rectified images outputted by \texttt{NSDRP} for each order \citep[see][for a description of optimal extraction]{Horne1986, Cushing2004}.  As an optimal profile, we used the slit profile returned by \texttt{NSDRP} for each order.  For each wavelength of each order, we fit the respective slit profile to the rectified image, allowing the normalization and additive offset of the profile to vary.  The offset accounts for any remaining background that was not removed by the AB subtraction.  After an initial fit, we removed data points that differed by more than 3$\sigma$ from the fitted profile, then refit the profile.  This approach rejected bad pixels (both ``hot'' pixels and ``dead'' pixels) from contaminating an entire wavelength bin.

Upon inspection of the resulting spectra, we found strong interference fringing effects with wavelength.  The fringing is attributed to the combination of the order-sorting filter and the long-wavelength blocking filter in the NIRSPEC-1 mode, neither of which was  replaced during the 2018 upgrade.  To remove the fringing, we followed the same approach as \citet[][]{Veyette2017}.  We Fourier transformed the spectrum of the featureless A0V star to find the frequencies of the fringes. We filtered the fringe frequencies in Fourier space with a FIR notch filter based on a Hanning window with a width of $6\times 10^{-3}$ pix$^{-1}$. Following this procedure, we inspected the resulting A0V spectrum and found no evidence of fringing.  We then applied an identical procedure to the \nirspectargets spectra.

\subsection{CARMENES Spectra}

CARMENES involves the combination of visible and near-infrared high-resolution spectrographs \citep[][]{Quirrenbach2016}.  The near-infrared spectrograph in CARMENES covers 9600 to 17100 \AA~with a resolving power of 80400.  \citet[][]{Reiners2018} describe publicly available spectra of 324 M dwarf stars from the CARMENES survey.  We downloaded the CARMENES near-infrared spectra from the CARMENES data archive.\footnote{\url{http://carmenes.cab.inta-csic.es/gto/welcome.action}}  For each CARMENES order within $Y$ band, we convolved the data with a Gaussian kernel corresponding to an R=25000 spectrograph.  Prior to convolution, the signal-to-noise of each publicly available spectrum was between 50 and 150 for each target.  The CARMENES data contained several gaps in wavelength coverage.  As we note below, we were unable to measure the equivalent width of one of the deep \ti lines due to these gaps. 

\citet{DiezAlonso2019} report rotational periods for 142 M dwarfs in the CARMENES survey, combining literature measurements with new photometric measurements.  We chose to analyze targets with literature rotational periods, as we considered those to be the most reliable.  \citet{Reiners2018} contains measurements of the \vsini of each spectrum, and as with the NIRSPEC data, we limited the sample to those with \vsini of less than 12 km s$^{-1}$.  We determined the masses and radii of the stars in the same manner as with the NIRSPEC sample.  Lastly, we visually inspected the resulting CARMENES spectra and removed one object owing to a significant amount of noise.  Following these cuts, we were left with \carmenestargets M dwarfs in the CARMENES sample.  Four of the \nirspectargets targets observed by NIRSPEC also have CARMENES data---HD 285968, G  99-49, V* YZ CMi, G 195-36---resulting in \totaltargets total objects with spectra in this work.  We used these objects to assess our uncertainties in measured equivalent widths.

Table \ref{tab:targets} lists the target properties and Figure \ref{fig:sample} shows the full target sample and the observed targets versus rotational period and Rossby number for all \totaltargets M dwarfs.   Convective turnover time was calculated using relations from \citet{Wright2011}.  The M dwarfs cover a reasonable spread of masses, rotational periods and Rossby numbers.

\begin{longrotatetable}
\begin{deluxetable*}{lcccccccccc}
\tablecaption{Details on M dwarf stars in this work\label{tab:targets}}
\tablehead{\colhead{SIMBAD Name} & 
\colhead{$\alpha$} & 
\colhead{$\delta$} & 
\colhead{$\mu$ (mas)} & 
\colhead{$\pi$ (mas)} & 
\colhead{$K$ mag} & 
\colhead{Mass ($M_\sun$)} & 
\colhead{Sp. Type} & 
\colhead{$P_{\rm rot}$ (days)} & 
\colhead{$\log Ro$} &
\colhead{Inst.}}
\startdata
BD+44  4548             & 00$^h$05$^m$10$^s$.89 & +45$^{\circ}$47$'$11$''.$64 & 870.75,-151.27 & 86.96 & 5.85 & 0.52 & M2Ve    & 15.37 & -0.41 & C \\
G  32-7                 & 00$^h$16$^m$16$^s$.15 & +19$^{\circ}$51$'$50$''.$47 & 708.09,-748.70 & 65.25 & 8.10 & 0.27 & M4.0V   & 105.00 & -0.31 & C \\
G  32-37                & 00$^h$39$^m$33$^s$.55 & +14$^{\circ}$54$'$19$''.$07 & 331.61,34.76 & 34.82 & 9.12 & 0.31 & M4.0V  & 34.00 & -0.65 & N \\
G  69-32                & 00$^h$54$^m$48$^s$.08 & +27$^{\circ}$31$'$03$''.$63 & 341.47,14.31 & 36.37 & 9.45 & 0.26 & M4.5V  & 1.70 & -2.13 & N \\
Wolf   47               & 01$^h$03$^m$19$^s$.83 & +62$^{\circ}$21$'$55$''.$83 & 730.74,86.35 & 101.64 & 7.72 & 0.20 & M5V     & 1.02 & -2.55 & C \\
YZ Cet                  & 01$^h$12$^m$30$^s$.60 & -16$^{\circ}$59$'$56$''.$00 & -4410.43,942.93 & 206.27 & 4.77 & 0.39 & M1.0Ve  & 69.20 & -0.15 & C \\
2MASS J01192628+5450382 & 01$^h$19$^m$26$^s$.28 & +54$^{\circ}$50$'$38$''.$23 & 166.39,-53.29 & 23.80 & 10.07 & 0.30 & M4.5V  & 17.60 & -0.99 & N \\
G 159-46                & 02$^h$12$^m$54$^s$.62 & +00$^{\circ}$00$'$16$''.$86 & 555.65,29.00 & 65.45 & 8.17 & 0.26 & M4V    & 4.70 & -1.69 & N \\
LP  197-37              & 02$^h$40$^m$52$^s$.42 & +44$^{\circ}$52$'$35$''.$07 & 279.20,90.21 & 45.25 & 8.46 & 0.33 & M4V    & 9.40 & -1.17 & N \\
LP  356-15              & 03$^h$24$^m$12$^s$.86 & +23$^{\circ}$46$'$19$''.$06 & 205.52,-110.35 & 48.45 & 7.45 & 0.46 & M2.5Ve & 20.60 & -0.52 & N \\
LP  413-24              & 03$^h$39$^m$07$^s$.13 & +20$^{\circ}$25$'$26$''.$71 & 186.45,-39.49 & 28.00 & 9.71 & 0.30 & M4.5   & 3.40 & -1.70 & N \\
G  80-21                & 03$^h$47$^m$23$^s$.34 & -01$^{\circ}$58$'$19$''.$95 & 180.67,-274.18 & 59.52 & 6.93 & 0.48 & M3.0V   & 3.87 & -1.23 & C \\
LP  357-119             & 04$^h$02$^m$24$^s$.42 & +24$^{\circ}$41$'$24$''.$42 & 146.50,-147.32 & 33.13 & 8.47 & 0.43 & M4.5   & 39.10 & -0.31 & N \\
GSC 05312-00079         & 04$^h$14$^m$17$^s$.31 & -09$^{\circ}$06$'$54$''.$61 & 99.82,-145.46 & 40.30 & 8.76 & 0.32 & M4.3V  & 1.70 & -1.93 & N \\
UCAC4 631-018323        & 04$^h$30$^m$18$^s$.23 & +36$^{\circ}$01$'$34$''.$26 & 193.54,-20.76 & 33.69 & 8.43 & 0.43 & M3.7V  & 36.40 & -0.34 & N \\
LP  834-32              & 04$^h$35$^m$36$^s$.19 & -25$^{\circ}$27$'$34$''.$59 & 67.29,-195.92 & 59.61 & 7.41 & 0.40 & M3.5   & 2.80 & -1.53 & N \\
HD 285968               & 04$^h$42$^m$55$^s$.78 & +18$^{\circ}$57$'$29$''.$39 & 656.38,-1116.50 & 105.56 & 5.61 & 0.49 & M2.5V  & 38.90 & -0.20 & B \\
RX J0451.0+3127         & 04$^h$51$^m$01$^s$.42 & +31$^{\circ}$27$'$23$''.$69 & 232.32,-47.48 & 48.73 & 8.16 & 0.35 & M3.7V  & 14.40 & -0.93 & N \\
G 100-46                & 05$^h$53$^m$22$^s$.97 & +22$^{\circ}$12$'$49$''.$78 & 166.70,-195.10 & 34.25 & 9.09 & 0.32 & M4V    & 19.50 & -0.87 & N \\
G  99-49                & 06$^h$00$^m$03$^s$.51 & +02$^{\circ}$42$'$23$''.$60 & 309.49,-40.64 & 192.07 & 6.04 & 0.23 & M3.5Ve  & 1.81 & -2.19 & B \\
G 192-15                & 06$^h$02$^m$29$^s$.19 & +49$^{\circ}$51$'$56$''.$16 & 67.22,-850.07 & 104.89 & 8.44 & 0.14 & M5.0V   & 105.00 & -0.90 & C \\
2MASS J06043887+0741545 & 06$^h$04$^m$38$^s$.87 & +07$^{\circ}$41$'$54$''.$55 & 84.37,-193.67 & 22.95 & 9.78 & 0.35 & M6V    & 7.30 & -1.22 & N \\
HD  42581               & 06$^h$10$^m$34$^s$.62 & -21$^{\circ}$51$'$52$''.$66 & -135.99,-719.09 & 173.70 & 4.17 & 0.55 & M1V     & 27.30 & -0.26 & C \\
UCAC4 533-032549        & 06$^h$44$^m$47$^s$.52 & +16$^{\circ}$28$'$17$''.$84 & 85.70,-139.76 & 26.03 & 9.24 & 0.39 & M3V    & 57.90 & -0.22 & N \\
UCAC4 686-047574        & 07$^h$10$^m$13$^s$.42 & +47$^{\circ}$00$'$58$''.$01 & -78.69,-170.24 & 26.70 & 8.89 & 0.44 & M3V    & 23.80 & -0.51 & N \\
LP  162-1               & 07$^h$17$^m$08$^s$.95 & +45$^{\circ}$45$'$54$''.$15 & 81.17,-148.37 & 28.13 & 9.48 & 0.33 & M4     & 115.80 & -0.08 & N \\
BD-02  2198             & 07$^h$36$^m$07$^s$.08 & -03$^{\circ}$06$'$38$''.$79 & 68.24,-289.99 & 71.08 & 5.93 & 0.57 & M1.0V   & 12.20 & -0.57 & C \\
UCAC4 480-038371        & 07$^h$37$^m$43$^s$.85 & +05$^{\circ}$54$'$36$''.$85 & -3.20,-153.64 & 27.26 & 9.08 & 0.40 & M3V    & 61.70 & -0.17 & N \\
UCAC3 229-91098         & 07$^h$38$^m$29$^s$.50 & +24$^{\circ}$00$'$08$''.$65 & -152.14,-96.86 & 51.59 & 8.12 & 0.33 & M2.7V  & 3.90 & -1.53 & N \\
V* YZ CMi               & 07$^h$44$^m$40$^s$.17 & +03$^{\circ}$33$'$08$''.$88 & -348.10,-445.88 & 167.02 & 5.70 & 0.32 & M4.0Ve & 2.80 & -1.73 & B \\
UCAC4 715-046733        & 07$^h$55$^m$12$^s$.08 & +52$^{\circ}$57$'$54$''.$01 & 148.50,-52.97 & 41.48 & 9.06 & 0.27 & M4     & 83.20 & -0.40 & N \\
G  50-21                & 08$^h$10$^m$53$^s$.62 & +03$^{\circ}$58$'$33$''.$61 & 114.28,-340.07 & 45.64 & 8.29 & 0.35 & M4V    & 123.40 & 0.00 & N \\
UCAC4 468-040412        & 08$^h$37$^m$30$^s$.22 & +03$^{\circ}$33$'$45$''.$90 & 61.72,-175.78 & 52.82 & 8.97 & 0.22 & M4.0V  & 100.90 & -0.50 & N \\
UCAC4 608-044702        & 08$^h$40$^m$15$^s$.99 & +31$^{\circ}$27$'$06$''.$81 & 205.54,120.00 & 89.21 & 7.30 & 0.28 & M3.5Ve  & 118.00 & -0.23 & C \\
G  46-27                & 09$^h$11$^m$12$^s$.70 & +01$^{\circ}$27$'$34$''.$87 & -20.36,-312.05 & 29.91 & 9.22 & 0.35 & M4V    & 120.10 & -0.01 & N \\
G 195-36                & 09$^h$42$^m$23$^s$.19 & +55$^{\circ}$59$'$01$''.$26 & -710.24,-508.33 & 60.44 & 7.53 & 0.37 & M3V    & 72.60 & -0.17 & B \\
BD+20  2465             & 10$^h$19$^m$36$^s$.28 & +19$^{\circ}$52$'$12$''.$02 & -498.61,-43.68 & 201.37 & 4.59 & 0.43 & M4Vae  & 2.20 & -1.57 & N \\
G 196-37                & 10$^h$36$^m$48$^s$.15 & +50$^{\circ}$55$'$04$''.$03 & 252.98,-206.70 & 41.85 & 9.02 & 0.27 & M4.5V  & 2.50 & -1.91 & N \\
DS Leo                  & 11$^h$02$^m$38$^s$.39 & +21$^{\circ}$58$'$01$''.$18 & -94.32,-671.25 & 58.91 & 6.55 & 0.54 & M1      & 14.60 & -0.55 & C \\
LP  263-64              & 11$^h$03$^m$09$^s$.99 & +36$^{\circ}$39$'$08$''.$60 & -188.38,30.09 & 43.81 & 8.63 & 0.31 & M3.5V  & 2.10 & -1.87 & N \\
K2-18                   & 11$^h$30$^m$14$^s$.52 & +07$^{\circ}$35$'$18$''.$26 & -80.38,-133.14 & 26.27 & 8.90 & 0.44 & M3.5Ve  & 36.40 & -0.29 & C \\
Ross 1003               & 11$^h$41$^m$44$^s$.64 & +42$^{\circ}$45$'$07$''.$10 & -575.65,-89.97 & 90.76 & 6.82 & 0.35 & M4.0Ve  & 71.50 & -0.23 & C \\
Ross  905               & 11$^h$42$^m$11$^s$.09 & +26$^{\circ}$42$'$23$''.$66 & 895.05,-814.03 & 102.50 & 6.07 & 0.42 & M3V     & 44.60 & -0.27 & C \\
G  10-49                & 11$^h$47$^m$40$^s$.75 & +00$^{\circ}$15$'$20$''.$10 & -314.20,-100.91 & 53.19 & 8.10 & 0.33 & M4V     & 11.60 & -1.08 & C \\
FI Vir                  & 11$^h$47$^m$44$^s$.36 & +00$^{\circ}$48$'$17$''.$14 & -326.00,-323.00 & 67.07 & 7.56 & 0.33 & M3V     & 163.00 & 0.09 & C \\
G 122-49                & 11$^h$50$^m$57$^s$.72 & +48$^{\circ}$22$'$38$''.$56 & -1545.70,-962.82 & 124.41 & 7.64 & 0.17 & M7      & 125.00 & -0.61 & C \\
Ross  689               & 12$^h$05$^m$29$^s$.68 & +69$^{\circ}$32$'$22$''.$60 & -456.97,-55.21 & 64.70 & 7.89 & 0.30 & M4V     & 100.00 & -0.25 & C \\
G 177-25                & 13$^h$10$^m$12$^s$.63 & +47$^{\circ}$45$'$18$''.$68 & -636.17,-616.24 & 82.03 & 8.69 & 0.16 & M5.0V   & 28.80 & -1.33 & C \\
NLTT 35712              & 13$^h$53$^m$38$^s$.80 & +77$^{\circ}$37$'$08$''.$14 & 224.14,-15.14 & 75.47 & 7.80 & 0.26 & M4.0V   & 1.23 & -2.24 & C \\
OT Ser                  & 15$^h$21$^m$52$^s$.88 & +20$^{\circ}$58$'$38$''.$79 & 2891.52,411.90 & 280.69 & 4.02 & 0.40 & M2V     & 3.37 & -1.44 & C \\
G 202-48                & 16$^h$25$^m$24$^s$.62 & +54$^{\circ}$18$'$14$''.$76 & 432.07,-171.67 & 154.47 & 5.83 & 0.32 & M1.5V   & 76.79 & -0.27 & C \\
BD-12  4523             & 16$^h$30$^m$18$^s$.06 & -12$^{\circ}$39$'$45$''.$32 & -94.03,-1183.78 & 232.21 & 5.07 & 0.30 & M3V     & 119.00 & -0.14 & C \\
G 204-39                & 17$^h$57$^m$50$^s$.97 & +46$^{\circ}$35$'$19$''.$13 & -15.18,578.30 & 71.48 & 7.00 & 0.40 & M2.5V   & 30.30 & -0.47 & C \\
LP  390-16              & 18$^h$13$^m$06$^s$.57 & +26$^{\circ}$01$'$51$''.$86 & 221.10,-37.19 & 56.01 & 8.06 & 0.32 & M4.0V   & 2.28 & -1.82 & C \\
BD+45  2743             & 18$^h$35$^m$18$^s$.39 & +45$^{\circ}$44$'$38$''.$54 & 452.19,363.67 & 64.25 & 6.08 & 0.58 & M0.5V   & 34.00 & -0.10 & C \\
Ross  149               & 18$^h$36$^m$19$^s$.23 & +13$^{\circ}$36$'$26$''.$37 & 179.14,278.91 & 83.03 & 7.37 & 0.29 & M4V     & 50.20 & -2.23 & C \\
G 141-36                & 18$^h$48$^m$17$^s$.53 & +07$^{\circ}$41$'$21$''.$18 & 376.33,249.30 & 131.20 & 7.91 & 0.14 & M5.0V   & 2.76 & -2.45 & C \\
Ross  154               & 18$^h$49$^m$49$^s$.37 & -23$^{\circ}$50$'$10$''.$45 & 639.35,-193.55 & 336.12 & 5.37 & 0.18 & M3.5Ve  & 2.87 & -2.23 & C \\
HD 176029               & 18$^h$58$^m$00$^s$.14 & +05$^{\circ}$54$'$29$''.$24 & -196.30,-1220.47 & 90.05 & 5.36 & 0.58 & M1.0Ve  & 35.20 & -0.08 & C \\
HD 180617               & 19$^h$16$^m$55$^s$.25 & +05$^{\circ}$10$'$08$''.$04 & -579.04,-1332.74 & 169.16 & 4.67 & 0.47 & M3-V    & 46.00 & -0.15 & C \\
G 185-18                & 19$^h$21$^m$38$^s$.70 & +20$^{\circ}$52$'$03$''.$27 & -948.34,-1455.73 & 94.20 & 7.93 & 0.20 & M4.0Ve  & 133.00 & -0.41 & C \\
Wolf 1069               & 20$^h$26$^m$05$^s$.30 & +58$^{\circ}$34$'$22$''.$68 & 261.08,542.99 & 104.32 & 8.10 & 0.16 & M4.95   & 57.70 & -0.61 & C \\
G 144-25                & 20$^h$40$^m$33$^s$.86 & +15$^{\circ}$29$'$58$''.$73 & 1321.00,662.23 & 104.89 & 7.75 & 0.19 & M4.5V   & 106.00 & -0.60 & C \\
LP  816-60              & 20$^h$52$^m$33$^s$.01 & -16$^{\circ}$58$'$29$''.$01 & -309.22,37.35 & 178.12 & 6.20 & 0.23 & M4V     & 67.60 & -0.62 & C \\
HD 209290               & 22$^h$02$^m$10$^s$.28 & +01$^{\circ}$24$'$00$''.$83 & -452.43,-278.58 & 94.74 & 5.32 & 0.57 & M0.5V   & 29.50 & -0.16 & C \\
EV Lac                  & 22$^h$46$^m$50$^s$.70 & +44$^{\circ}$20$'$08$''.$00 & -483.17,89.27 & 92.90 & 6.07 & 0.46 & M3.0Ve  & 4.38 & -1.20 & C \\
IL Aqr                  & 22$^h$53$^m$16$^s$.50 & -14$^{\circ}$15$'$48$''.$00 & -94.32,-671.25 & 58.91 & 6.55 & 0.54 & M1      & 81.00 & 0.29 & C \\
HD 216899               & 22$^h$56$^m$34$^s$.80 & +16$^{\circ}$33$'$12$''.$36 & -1034.80,-284.00 & 145.61 & 4.52 & 0.55 & M1.5Ve  & 39.50 & -0.11 & C \\
HH And                  & 23$^h$41$^m$54$^s$.00 & +44$^{\circ}$10$'$46$''.$00 & 748.11,480.60 & 190.26 & 5.95 & 0.24 & M4.0Ve  & 106.00 & -0.41 & C \\
RX J2354.8+3831         & 23$^h$54$^m$51$^s$.46 & +38$^{\circ}$31$'$36$''.$20 & -131.57,-86.11 & 59.35 & 8.09 & 0.30 & M3.1V   & 4.70 & -1.56 & C \\
\enddata
\tablecomments{All astrometric data are from Gaia DR2 in ICRS and epoch 2000 \citep{Gaia2018}.  $K$-band magnitudes are from the Two Micron All Sky Survey Point Source Catalog \citep[2MASS,][]{Cutri2003,Skrutskie2006}.  Masses were calculated using using $M_K$-to-mass relations of \citet{Mann2019}.  Spectral types from SIMBAD.  Rotation periods are from \citet{Newton2017} for targets with NIRSPEC observations and from \citet{DiezAlonso2019} for CARMENES data.  For targets with both, we default to the \citet{Newton2017} rotation periods.  Convective turnover time was calculated using the mass-dependent relation from \citet{Wright2011}. ``C'' refers to CARMENES, ``N'' refers to NIRSPEC and ``B'' indicates both NIRSPEC and CARMENES.  A machine-readable version of this table is available in online version of the manuscript.}
\end{deluxetable*}
\end{longrotatetable}

\section{Analysis}\label{sec:analysis}

For the combined \totaltargets M dwarfs, we measured the equivalent widths of 10 Ti I lines originally identified by \citet{Veyette2017} as tracers of Ti abundance in M dwarf stars.  We used the \texttt{analyze\_NIRSPEC1} software pipeline to measure equivalent widths, available on GitHub.  See \citet{Veyette2017} for a detailed description of the method for measuring equivalent widths, which we briefly summarize here.  Spectra of M dwarf stars contain significant molecular features across all wavelengths, which result in a pseudocontinuum useful for measuring equivalent widths.  In $Y$ band, the dominant source of molecular opacity is FeH.  

First, we cross-correlated each spectrum with synthetic spectra from the BT-Settl atmosphere database \citep[][]{Allard2012} in order to remove any radial velocity shifts.  To estimate the pseudocontinuum, first we removed high-frequency variations in the spectrum with a second-order Savitzky-Golay filter with a window length of five pixels \citep[][]{Savitzky1964}.  We then applied a running maximum filter with a width of seven resolution elements to establish a ceiling to the features. Lastly, we fitted a sixth-order Chebyshev polynomial to the filtered spectrum, which was taken as the pseudocontinuum for measuring equivalent widths.

Of the 10 \ti lines identified by \citet{Veyette2017}, we report on the six deepest lines for the NIRSPEC data and five lines for the CARMENES data.  Both the NIRSPEC and CARMENES data also include a deep \ca line, which we also report.  Table \ref{tab:lines} displays the properties of these six lines, which we refer to as \ti (1) to \ti (6).  The CARMENES data is missing the \ti (5) line due to gaps in the publicly available spectra near the edge of the line.  

The line properties were retrieved from the Third Vienna Atomic Line Database \citep[VALD3][]{Piskunov1995, Ryabchikova2015}, and the line data originates from \citet{K10}.  All of the \ti lines have similar lower energy states, but with a range of absorption cross-sections, parameterized as the log of the transition oscillator strength (f) times the statistical weight (g), and effective \lande g factors.

\begin{deluxetable*}{ccccccc}[t!]
\tablecaption{\ti Absorption Line Properties \label{tab:lines}}
\tablecolumns{7}
\tablewidth{0pt}
\tablehead{
\colhead{Line} &
\colhead{Wavelength in Air (\AA)} &
\colhead{Lower Energy (wn)} & 
\colhead{Lower State} & 
\colhead{Upper State} &
\colhead{$\log gf$} &
\colhead{\lande $g_{\rm eff}$}
}
\startdata
\ca & 10343.8194 & 23652.3040 & 3p6.4s.4p 1P* &  3p6.4s.5s 1S & -0.300 &  1.00 \\
\ti (1) & 10396.802 & 6842.965 & 3d3.(2G).4s a3G & 3d2.(3F).4s.4p.(1P*) y3F* & -1.54 & 1.13\\
\ti (2) & 10496.113 & 6742.755 & 3d3.(4F).4s b3F & 3d2.(3F).4s.4p.(3P) z3G* & -1.65 & 1.05\\
\ti (3) & 10584.633 & 6661.004 & 3d3.(2D2).4s a3D & 3d2.(1D).4s.4p.(3P*) x3D* & -1.77 & 1.00\\
\ti (4) & 10677.047 & 6742.755 & 3d3.(4F).4s a5F & 3d2.(3F).4s.4p.(3P*) z5G* & -2.52 & 1.25\\
\ti (5) & 10726.391 & 6556.833 & 3d3.(4F).4s a5F & 3d2.(3F).4s.4p.(3P*) z5G* & -2.06 & 0.59\\
\ti (6) & 10774.866 & 6598.764 & 3d3.(4F).4s a5F & 3d2.(3F).4s.4p.(3P*) z5G* & -2.67 & 0.69\\
\enddata
\tablecomments{Line properties were accessed from the VALD3 database \citep[][]{Piskunov1995, Ryabchikova2015}.  Line data originates from \citet{K10}.}
\end{deluxetable*}

\begin{figure}
    \centering
    \includegraphics[width=0.49\textwidth]{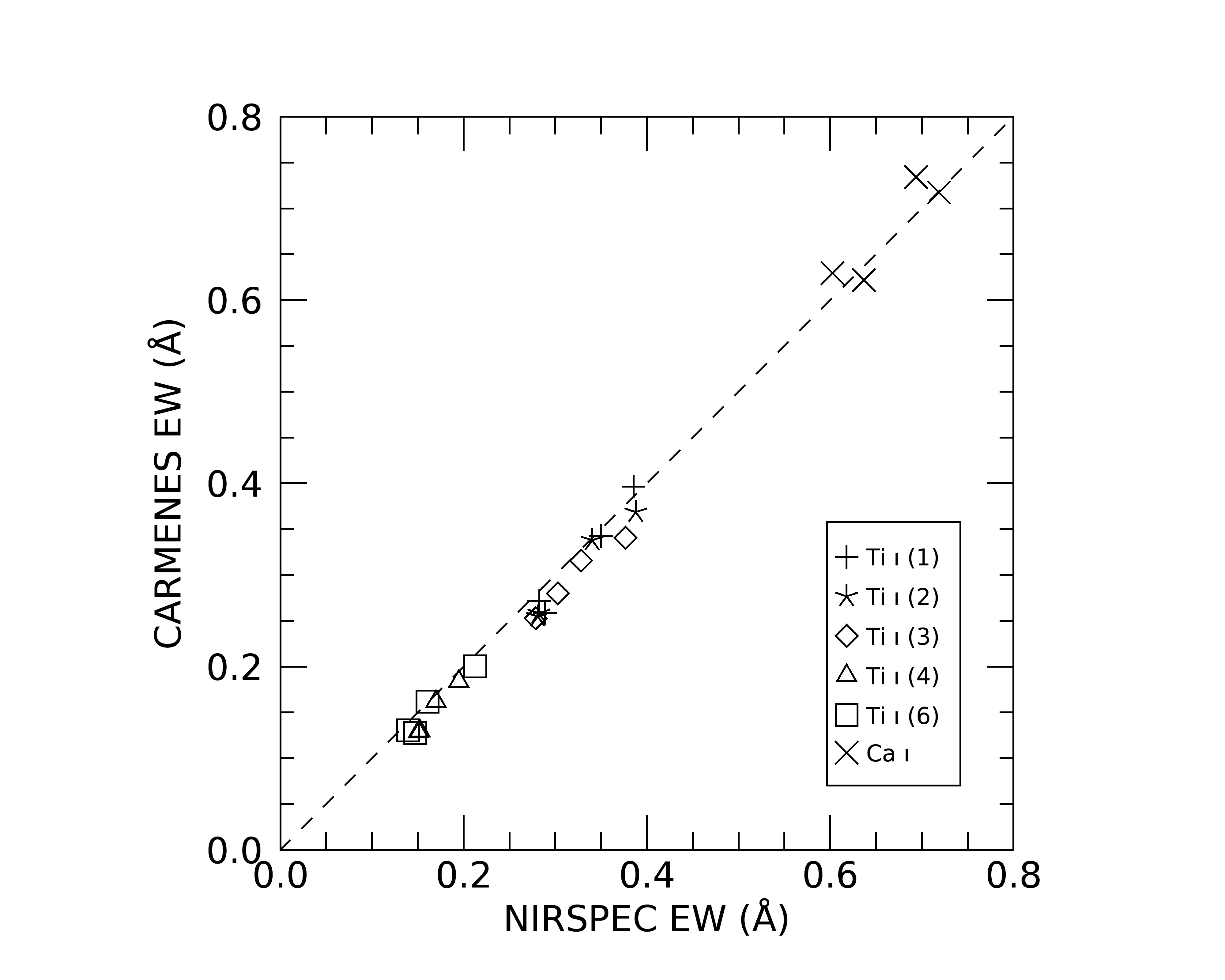}
    \caption{Equivalent widths of \ti and \ca lines in $Y$ band measured using NIRSPEC data versus CARMENES data on the same stars.  The dashed line represents a one-to-one correspondance.  The CARMENES equivalent widths are systematically lower than the NIRSPEC equivalent widths by  0.011 \AA, and the standard deviation of the differences is 0.018 \AA.}
    \label{fig:compare_ews}
\end{figure}

For the four stars with both NIRSPEC and CARMENES data, we used the differences in the equivalent width determinations as a rubric for uncertainties.  Figure \ref{fig:compare_ews} plots the equivalents widths measured from NIRSPEC data versus those measured from CARMENES data for the same stars.   The CARMENES equivalent widths are systematically lower than the NIRSPEC equivalent widths by 0.011 \AA, and the standard deviation of the differences is 0.018 \AA.  Since the systematic difference is within the standard deviation, we do not apply any corrections to the CARMENES equivalent widths.  We adopt 0.018 \AA~ as the uncertainty in all equivalent width measurements.  We adopt this value as a conservative estimate of the typical uncertainties, as it incorporates systematic errors in the equivalent width calculation that are difficult to include via canonical error propagation.  All of the spectra have similar signal-to-noise values and we do not expect much variation in the equivalent width uncertainties between objects. 

\figsetstart
\figsetnum{3}
\figsettitle{$Y$-Band Spectra}

\figsetgrpstart
\figsetgrpnum{3.3}
\figsetgrptitle{NIRSPEC $Y$ band spectra of G  32-37                taken as part of this work.  \textit{Top}: Full $Y$-band spectrum continuum-normalized and shifted to zero velocity with features indicated.  \textit{Bottom}: Close up on the six \ion{Ti}{I} lines used in this analysis, with the transition wavelength (dashed line) and limits of the equivalent width calculation (dotted lines) indicated.}
\figsetplot{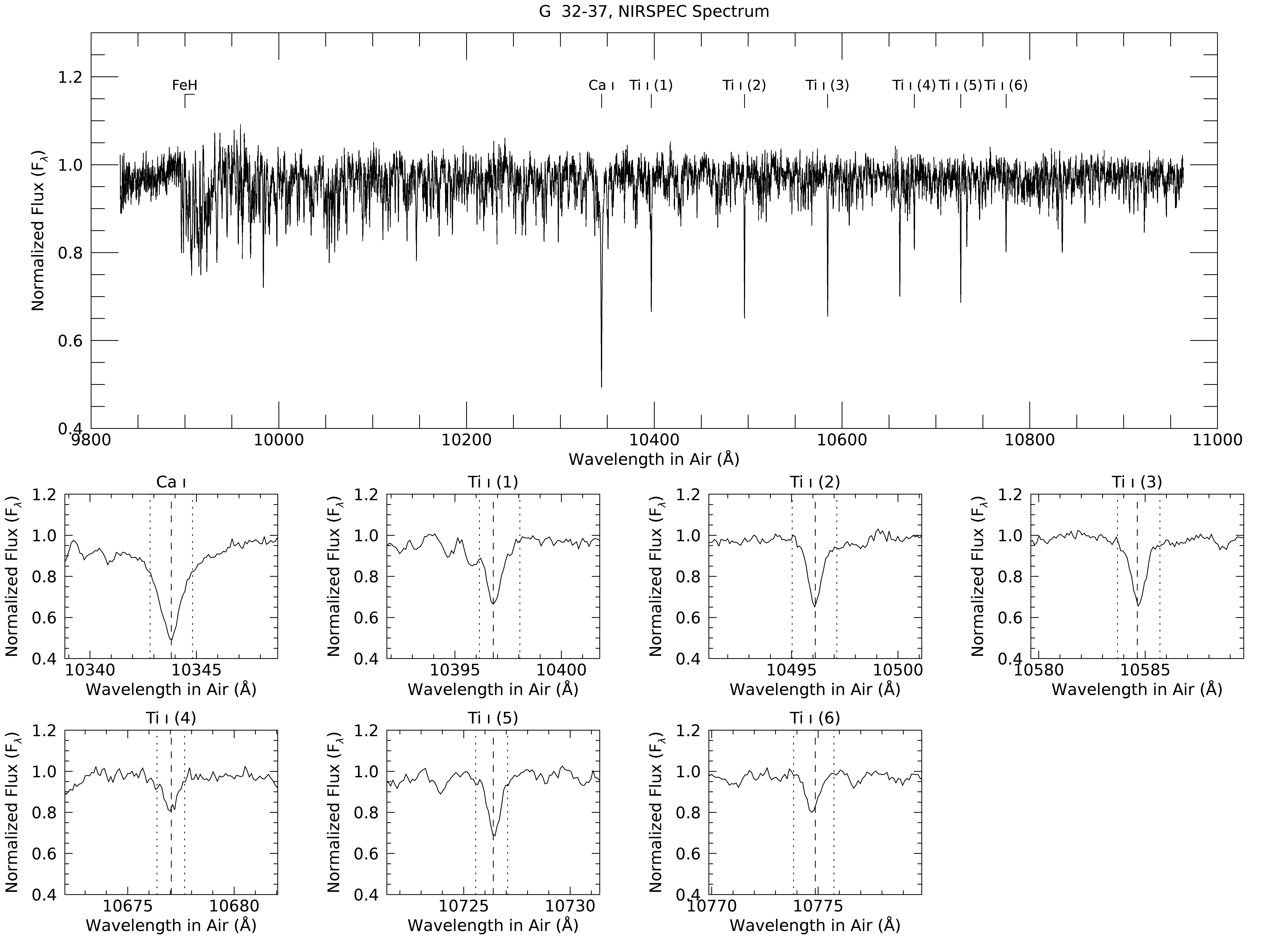}
\figsetgrpnote{\textit{Top}: Full $Y$-band spectrum continuum-normalized and shifted to zero velocity with features indicated.  \textit{Bottom}: Close up on the \ti lines used in this analysis, with the VALD transition wavelength (dashed line) and limits of the equivalent width calculation (dotted lines) indicated.  Figures and data for all objects in the sample are included in a figure set available in the online journal.}
\figsetgrpend

\figsetend

\begin{figure*}
\includegraphics[width=0.99\textwidth]{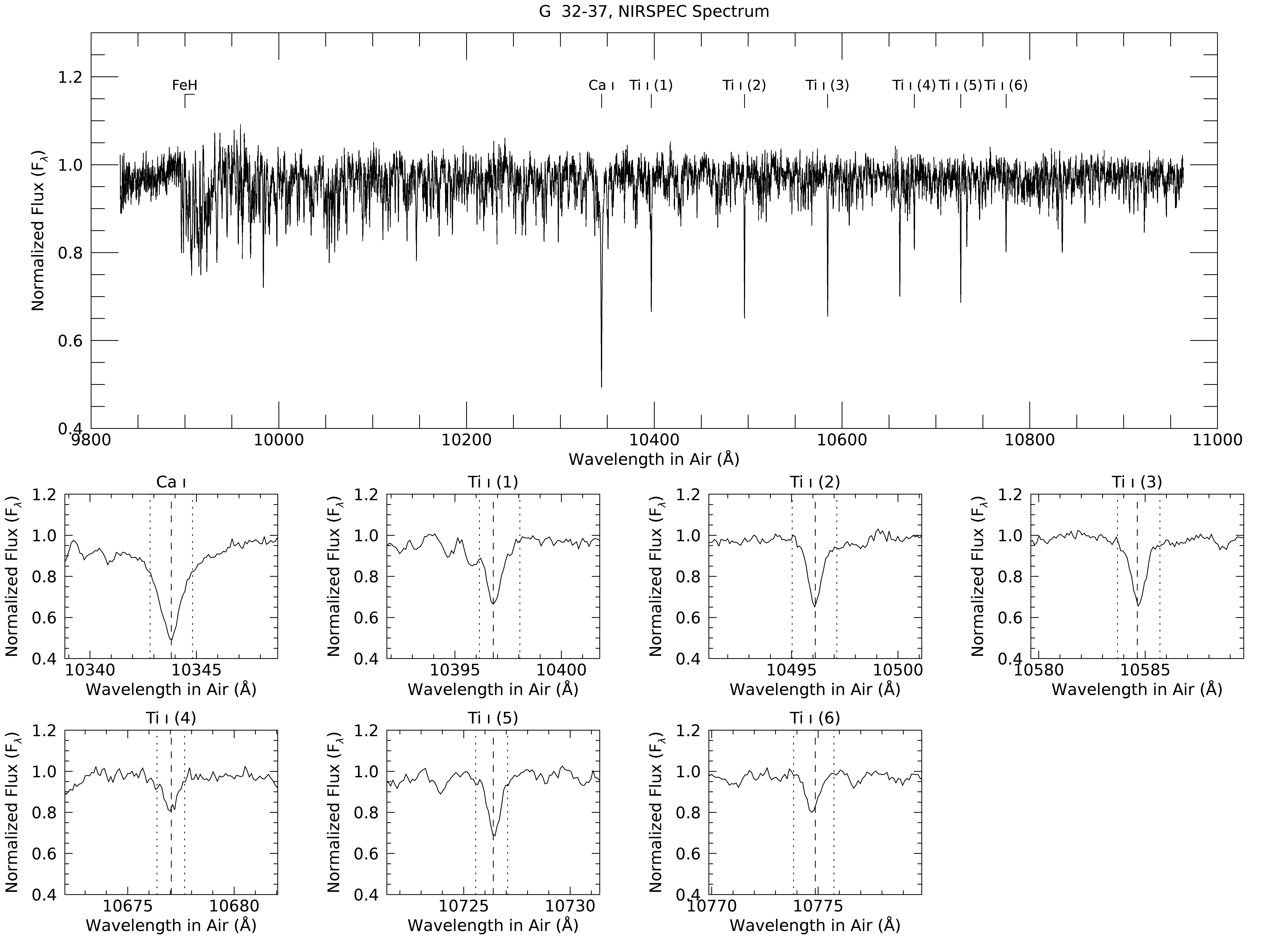}
\caption{\textit{Top}: Full $Y$-band spectrum continuum-normalized and shifted to zero velocity with features indicated.  \textit{Bottom}: Close up on the \ti lines used in this analysis, with the VALD transition wavelength (dashed line) and limits of the equivalent width calculation (dotted lines) indicated.  Figures and data for all objects in the sample are included in a figure set available in the online journal.}
\label{fig:spectra}
\end{figure*}

\begin{figure*}[]
\includegraphics[width=0.99\textwidth]{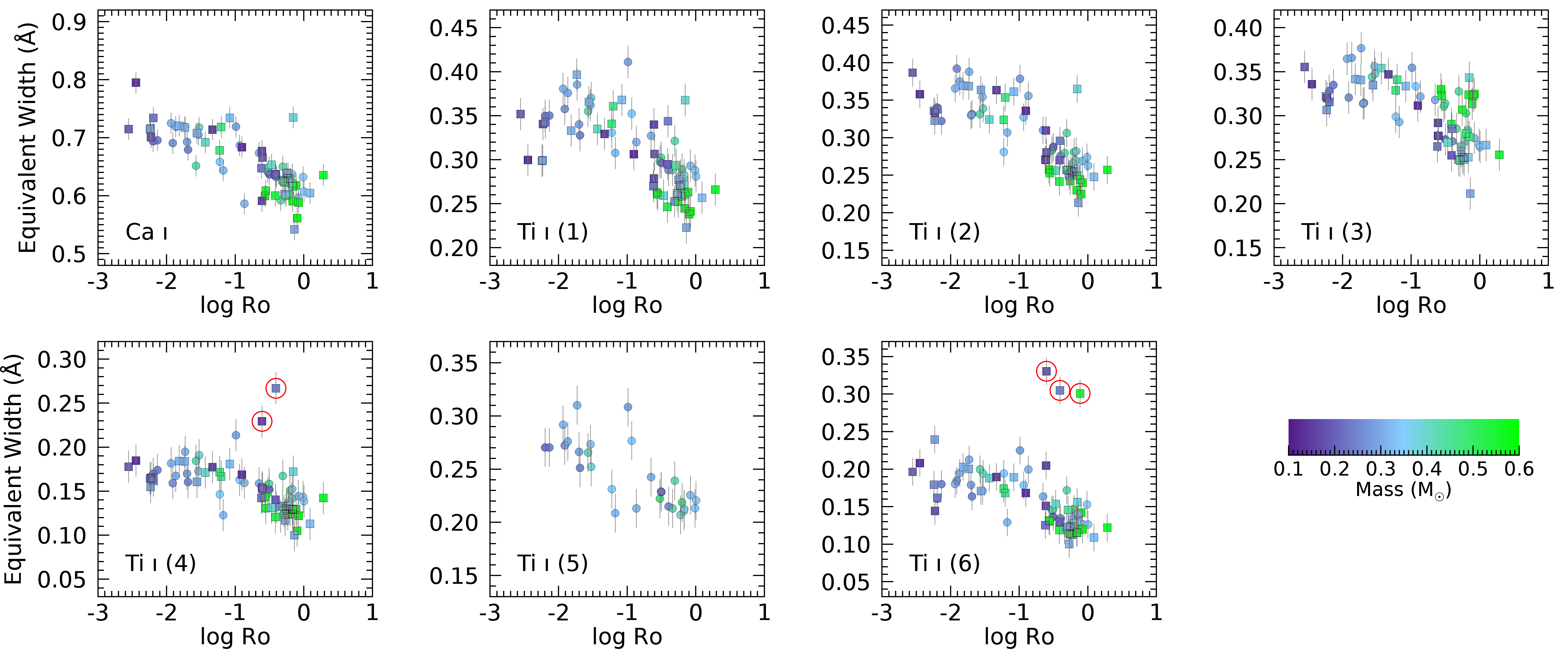}
\caption{Equivalent width versus Rossby number ($Ro$) for the \ca and \ti lines of the \totaltargets M dwarfs in the combined NIRSPEC and CARMENES samples. M dwarfs with NIRSPEC data are shown as circles; M dwarfs with CARMENES data are shown as squares.  We lack CARMENES data for the \ti (5) line.  The colorbar indicates the M dwarf mass, determined using mass-luminosity relationships from \citet{Mann2019}.  The outliers in \ti (4) and \ti (6) appear to be due to contaminated CARMENES spectra and are marked with a red circle.\label{fig:ew_v_ro}}
\end{figure*}

\begin{figure*}[]
\includegraphics[width=0.99\textwidth]{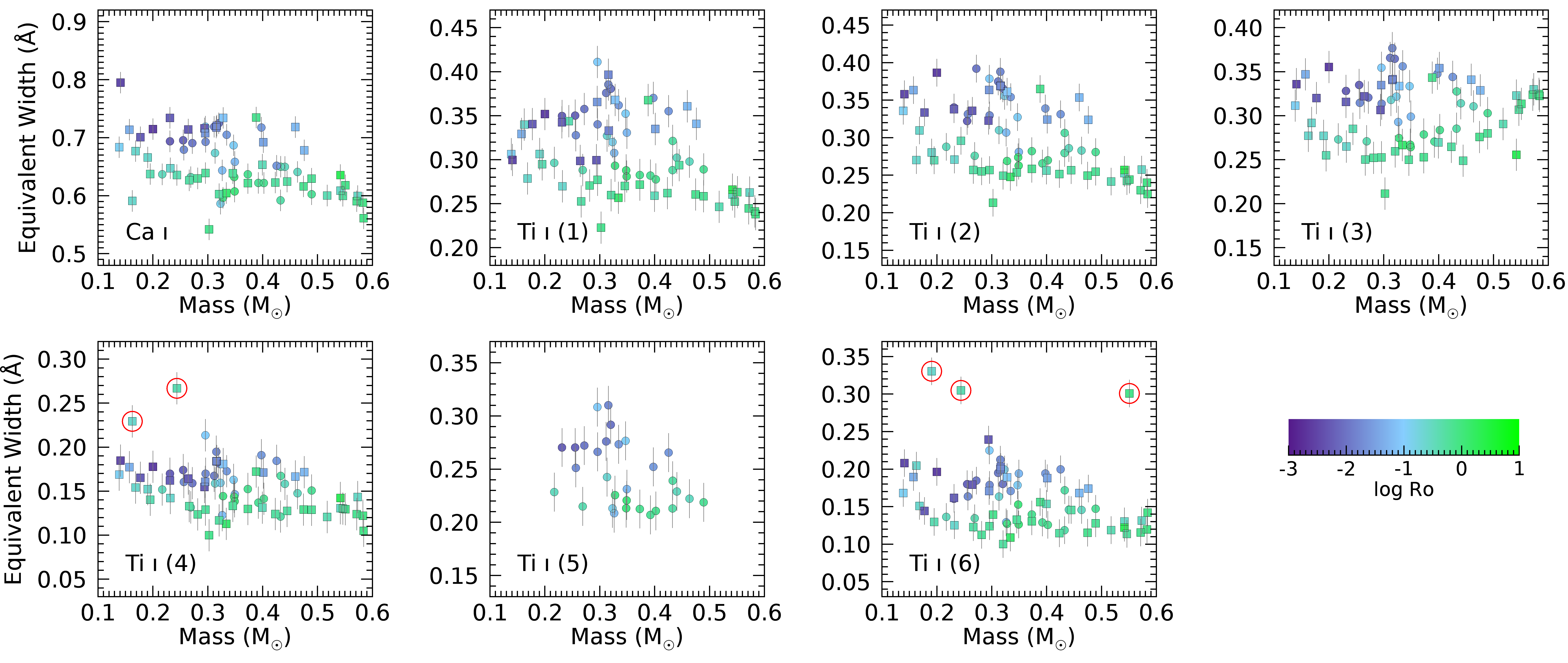}
\caption{Same as Figure \ref{fig:ew_v_ro}, but with stellar mass on the x-axis and colored by Rossby number.  There is some mass/temperature dependence to the equivalent widths; however, at fixed mass, we see an increase in equivalent width with Rossby number.\label{fig:ew_v_mass}}
\end{figure*}

Figure \ref{fig:spectra} and the accompanying online figure set show the resulting continuum-normalized and zero-velocity spectra as well as the  lines used in this work.  Table \ref{tab:results} lists the resulting equivalent widths for the six \ti lines and one \ca line.

\begin{longrotatetable}
\begin{deluxetable*}{lcccccccr}
\tablecaption{\ti Equivalent Widths (\AA) \label{tab:results}}
\tablehead{\colhead{SIMBAD Name} & 
\colhead{\ti (1)} & 
\colhead{\ti (2)} & 
\colhead{\ti (3)} & 
\colhead{\ti (4)} & 
\colhead{\ti (5)} & 
\colhead{\ti (6)} &
\colhead{\ca} &
\colhead{NIRSPEC/CARMENES}}
\startdata
BD+44  4548             & 0.2464 & 0.2415 & 0.2905 & 0.1206 &  & 0.1187 & 0.6002 & CARMENES\\
G  32-7                 & 0.2526 & 0.2570 & 0.2502 & 0.1329 &  & 0.1228 & 0.6265 & CARMENES\\
G  32-37                & 0.3272 & 0.3099 & 0.3179 & 0.1587 & 0.2425 & 0.1634 & 0.6735 & NIRSPEC\\
G  69-32                & 0.3501 & 0.3221 & 0.3349 & 0.1739 & 0.2703 & 0.1799 & 0.6954 & NIRSPEC\\
Wolf   47               & 0.3519 & 0.3865 & 0.3553 & 0.1777 &  & 0.1963 & 0.7147 & CARMENES\\
V* YZ Cet               & 0.3677 & 0.3649 & 0.3432 & 0.1721 &  & 0.1560 & 0.7348 & CARMENES\\
2MASS J01192628+5450382 & 0.4110 & 0.3785 & 0.3544 & 0.2135 & 0.3083 & 0.2249 & 0.7190 & NIRSPEC\\
G 159-46                & 0.3277 & 0.3313 & 0.3146 & 0.1604 & 0.2511 & 0.1636 & 0.6793 & NIRSPEC\\
LP  197-37              & 0.3076 & 0.3068 & 0.2928 & 0.1226 & 0.2085 & 0.1291 & 0.6435 & NIRSPEC\\
LP  356-15              & 0.2978 & 0.2828 & 0.3106 & 0.1476 & 0.2221 & 0.1456 & 0.6408 & NIRSPEC\\
LP  413-24              & 0.3401 & 0.3297 & 0.3136 & 0.1698 & 0.2662 & 0.1788 & 0.6928 & NIRSPEC\\
G  80-21                & 0.3408 & 0.3237 & 0.3287 & 0.1716 &  & 0.1740 & 0.6781 & CARMENES\\
LP  357-119             & 0.3212 & 0.3062 & 0.3275 & 0.1673 & 0.2391 & 0.1718 & 0.6499 & NIRSPEC\\
GSC 05312-00079         & 0.3805 & 0.3656 & 0.3646 & 0.1815 & 0.2917 & 0.1802 & 0.7248 & NIRSPEC\\
UCAC4 631-018323        & 0.2881 & 0.2795 & 0.2852 & 0.1210 & 0.2130 & 0.1186 & 0.5919 & NIRSPEC\\
LP  834-32              & 0.3701 & 0.3388 & 0.3475 & 0.1908 & 0.2521 & 0.1942 & 0.7175 & NIRSPEC\\
HD 285968               & 0.2889 & 0.2808 & 0.3028 & 0.1506 & 0.2187 & 0.1472 & 0.6024 & NIRSPEC\\
HD 285968               & 0.2584 & 0.2547 & 0.2798 & 0.1288 &  & 0.1277 & 0.6294 & CARMENES\\
RX J0451.0+3127         & 0.3523 & 0.3273 & 0.3335 & 0.1628 & 0.2766 & 0.1786 & 0.6868 & NIRSPEC\\
G 100-46                & 0.3200 & 0.3557 & 0.3216 & 0.1593 & 0.2130 & 0.1994 & 0.5862 & NIRSPEC\\
G  99-49                & 0.3424 & 0.3378 & 0.3157 & 0.1620 &  & 0.1617 & 0.7342 & CARMENES\\
G  99-49                & 0.3497 & 0.3402 & 0.3281 & 0.1698 & 0.2703 & 0.1605 & 0.6937 & NIRSPEC\\
G 192-15                & 0.3063 & 0.3357 & 0.3114 & 0.1688 &  & 0.1681 & 0.6836 & CARMENES\\
2MASS J06043887+0741545 & 0.3306 & 0.2809 & 0.2989 & 0.1464 & 0.2312 & 0.1941 & 0.6583 & NIRSPEC\\
HD  42581               & 0.2524 & 0.2422 & 0.3067 & 0.1301 &  & 0.1137 & 0.5997 & CARMENES\\
UCAC4 533-032549        & 0.2820 & 0.2656 & 0.2705 & 0.1369 & 0.2069 & 0.1289 & 0.6219 & NIRSPEC\\
UCAC4 686-047574        & 0.3022 & 0.2859 & 0.3140 & 0.1582 & 0.2289 & 0.1459 & 0.6497 & NIRSPEC\\
LP  162-1               & 0.2931 & 0.2689 & 0.2745 & 0.1443 & 0.2255 & 0.1271 & 0.5964 & NIRSPEC\\
BD-02  2198             & 0.2627 & 0.2574 & 0.3299 & 0.1433 &  & 0.1315 & 0.5995 & CARMENES\\
UCAC4 480-038371        & 0.2776 & 0.2698 & 0.2836 & 0.1413 & 0.2107 & 0.1257 & 0.6218 & NIRSPEC\\
UCAC3 229-91098         & 0.3619 & 0.3541 & 0.3561 & 0.1727 & 0.2733 & 0.1708 & 0.7049 & NIRSPEC\\
V* YZ CMi               & 0.3854 & 0.3878 & 0.3767 & 0.1947 & 0.3101 & 0.2126 & 0.7187 & NIRSPEC\\
V* YZ CMi               & 0.3965 & 0.3686 & 0.3405 & 0.1839 &  & 0.2002 & 0.7174 & CARMENES\\
UCAC4 715-046733        & 0.2881 & 0.2758 & 0.2708 & 0.1315 & 0.2149 & 0.1345 & 0.6319 & NIRSPEC\\
G  50-21                & 0.2809 & 0.2626 & 0.2642 & 0.1385 & 0.2206 & 0.1263 & 0.6072 & NIRSPEC\\
UCAC4 468-040412        & 0.2963 & 0.2879 & 0.2730 & 0.1518 & 0.2284 & 0.1363 & 0.6368 & NIRSPEC\\
UCAC4 608-044702        & 0.2706 & 0.2551 & 0.2521 & 0.1236 &  & 0.1125 & 0.6298 & CARMENES\\
G  46-27                & 0.2879 & 0.2738 & 0.2670 & 0.1434 & 0.2132 & 0.1528 & 0.6321 & NIRSPEC\\
G 195-36                & 0.2825 & 0.2820 & 0.2786 & 0.1523 & 0.2125 & 0.1396 & 0.6368 & NIRSPEC\\
G 195-36                & 0.2717 & 0.2584 & 0.2528 & 0.1298 &  & 0.1305 & 0.6216 & CARMENES\\
BD+20  2465             & 0.3551 & 0.3312 & 0.3440 & 0.1844 & 0.2655 & 0.1997 & 0.6515 & NIRSPEC\\
G 196-37                & 0.3575 & 0.3919 & 0.3206 & 0.1590 & 0.2722 & 0.1846 & 0.6906 & NIRSPEC\\
V* DS Leo               & 0.2604 & 0.2528 & 0.3228 & 0.1306 &  & 0.1303 & 0.6086 & CARMENES\\
LP  263-64              & 0.3758 & 0.3748 & 0.3656 & 0.1673 & 0.2760 & 0.1949 & 0.7187 & NIRSPEC\\
K2-18                   & 0.2937 & 0.2566 & 0.2488 & 0.1275 &  & 0.1455 & 0.6243 & CARMENES\\
Ross 1003               & 0.2701 & 0.2536 & 0.2499 & 0.1332 &  & 0.1326 & 0.6387 & CARMENES\\
Ross  905               & 0.2620 & 0.2514 & 0.2646 & 0.1240 &  & 0.1146 & 0.6226 & CARMENES\\
G  10-49                & 0.3679 & 0.3612 & 0.3336 & 0.1807 &  & 0.1891 & 0.7340 & CARMENES\\
Ross  128               & 0.2567 & 0.2477 & 0.2666 & 0.1128 &  & 0.1088 & 0.6045 & CARMENES\\
G 122-49                & 0.2786 & 0.3097 & 0.2918 & 0.1541 &  & 0.1509 & 0.6768 & CARMENES\\
Ross  689               & 0.2771 & 0.2568 & 0.2526 & 0.1290 &  & 0.1239 & 0.6390 & CARMENES\\
G 177-25                & 0.3292 & 0.3633 & 0.3470 & 0.1772 &  & 0.1894 & 0.7136 & CARMENES\\
NLTT 35712              & 0.2986 & 0.3359 & 0.3221 & 0.1642 &  & 0.1793 & 0.7141 & CARMENES\\
V* OT Ser               & 0.3349 & 0.3241 & 0.3541 & 0.1711 &  & 0.1880 & 0.6921 & CARMENES\\
G 202-48                & 0.2597 & 0.2494 & 0.2594 & 0.1168 &  & 0.1001 & 0.6026 & CARMENES\\
BD-12  4523             & 0.2228 & 0.2133 & 0.2114 & 0.0999 &  & 0.1393 & 0.5419 & CARMENES\\
G 204-39                & 0.2590 & 0.2562 & 0.2696 & 0.1313 &  & 0.1537 & 0.6534 & CARMENES\\
LP  390-16              & 0.3330 & 0.3692 & 0.3415 & 0.1841 &  & 0.2028 & 0.7206 & CARMENES\\
BD+45  2743             & 0.2381 & 0.2248 & 0.3225 & 0.1049 &  & 0.1420 & 0.5610 & CARMENES\\
Ross  149               & 0.2994 & 0.3226 & 0.3065 & 0.1546 &  & 0.2394 & 0.7162 & CARMENES\\
G 141-36                & 0.2996 & 0.3579 & 0.3357 & 0.1847 &  & 0.2080 & 0.7949 & CARMENES\\
Ross  154               & 0.3404 & 0.3334 & 0.3200 & 0.1651 &  & 0.1443 & 0.7007 & CARMENES\\
HD 176029               & 0.2413 & 0.2401 & 0.3247 & 0.1221 &  & 0.1197 & 0.5879 & CARMENES\\
HD 180617               & 0.2605 & 0.2494 & 0.2758 & 0.1290 &  & 0.1152 & 0.6160 & CARMENES\\
G 185-18                & 0.2947 & 0.2699 & 0.2550 & 0.1402 &  & 0.1296 & 0.6373 & CARMENES\\
Wolf 1069               & 0.3399 & 0.2701 & 0.2771 & 0.2293 &  & 0.2046 & 0.5909 & CARMENES\\
G 144-25                & 0.3064 & 0.2802 & 0.2771 & 0.1522 &  & 0.3304 & 0.6658 & CARMENES\\
LP  816-60              & 0.2698 & 0.2707 & 0.2649 & 0.1421 &  & 0.1252 & 0.6472 & CARMENES\\
HD 209290               & 0.2446 & 0.2300 & 0.3237 & 0.1238 &  & 0.1156 & 0.5904 & CARMENES\\
V* EV Lac               & 0.3607 & 0.3535 & 0.3408 & 0.1666 &  & 0.1681 & 0.7183 & CARMENES\\
BD-15  6290             & 0.2659 & 0.2570 & 0.2555 & 0.1422 &  & 0.1220 & 0.6355 & CARMENES\\
HD 216899               & 0.2630 & 0.2441 & 0.3134 & 0.1294 &  & 0.3008 & 0.6179 & CARMENES\\
Ross  248               & 0.3437 & 0.2958 & 0.2850 & 0.2668 &  & 0.3048 & 0.6356 & CARMENES\\
RX J2354.8+3831         & 0.3655 & 0.3636 & 0.3347 & 0.1606 &  & 0.1708 & 0.7077 & CARMENES\\
\enddata
\tablecomments{A machine-readable version of this table is available in the online version of the manuscript.}
\end{deluxetable*}
\end{longrotatetable}

Figure \ref{fig:ew_v_ro} clearly show an increase in \ti equivalent widths with decreasing Rossby number despite the appearance of several outlying measurements.  Visual inspection suggests that the equivalent widths saturate for $Ro \lesssim 0.1$, as is seen in H$\alpha$, X-ray emission and FeH broadening, for most of the lines analyzed.  We attribute the trend and saturation to magnetic enhancement of the lines.  For deep absorption lines with saturated line cores (saturation here referring to the physical saturation of the line core, not the saturation of equivalent width with Rossby number), Zeeman splitting can dramatically increase the line equivalent width, even if the total number of absorbing atoms remains the same.   This effect occurs prior to stellar rotational broadening and instrumental broadening, so the effect of increased equivalent width can occur in lines that do not appear saturated in measured spectra, even if they have saturated line cores, as is the case here.

Figure \ref{fig:ew_v_mass} is identical to Figure \ref{fig:ew_v_ro}, except with mass on the x-axis and points colored by Rossby number.  We see some equivalent width dependence on stellar mass (and correspondingly effective temperature); however, at fixed mass, the increase and saturation with equivalent width with Rossby number is evident, especially around 0.3 $M_\sun$.

We note that several objects appear as outliers in \ref{fig:ew_v_ro}.  For \ti (6), the CARMENES spectra of G 144-25, HD 216899 and Ross 248 appear to be contaminated by bad pixels, leading to erroneously large equivalent width determinations.  Similarly, for \ti (4), the CARMENES spectra of Wolf 1069 and Ross 248 appear to be contaminated by bad pixels, resulting in erroneous equivalent width determinations.

\subsection{The Curve of Growth}

\begin{figure}[]
\includegraphics[width=0.49\textwidth]{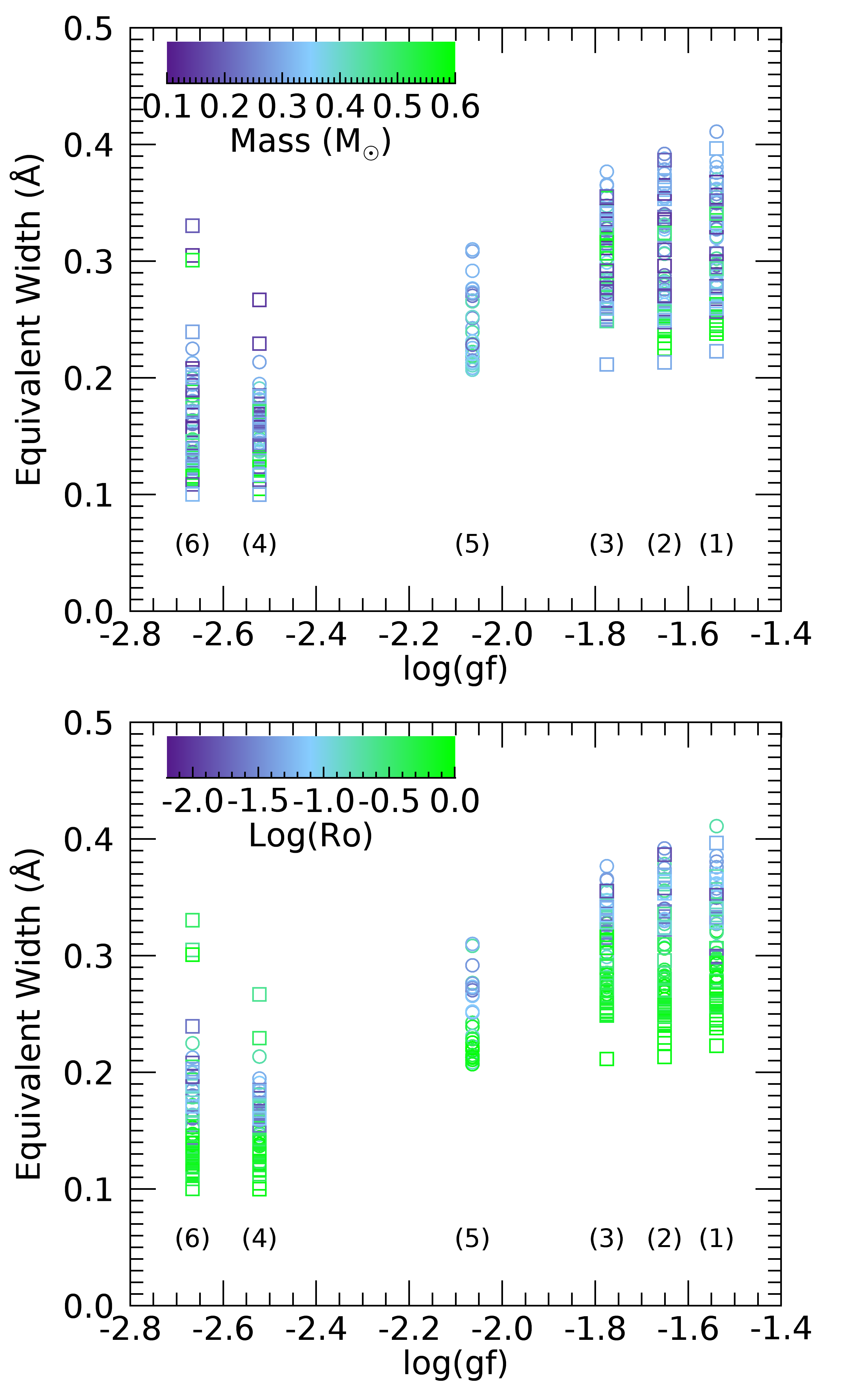}
\caption{Equivalent width versus the log of the oscillator strength $f$ times the number of degenerate states $g$ (the ``curve of growth'') for the six \ti lines in this study, colored by stellar mass (\textit{top}) and by log of the Rossby number (\textit{bottom}).  Measurements from NIRSPEC data are circles and measurements from CARMENES data are squares.  The sub-linear increase in equivalent width with $gf$ suggests the lines are in the saturated or damping regime and are therefor subject to magnetic enhancement.\label{fig:cog}}
\end{figure}

 Since the \ti lines originate from the same atomic species and ionization state, and the energies of the lower and upper states are similar, they can be plotted on a ``curve-of-growth'' diagram, showing equivalent width versus $\log (gf)$.  Figure \ref{fig:cog} shows the resulting curve-of-growth diagrams for the lines, colored by stellar mass and by Rossby number ($Ro$).  Whereas the curves of growth do not show a clear dependence on stellar mass, they do show a dependence on Rossby number.
 
The relatively small increase in equivalent width versus $gf$ (a factor of two in equivalent width versus a decade in $gf$) suggests that all six lines are located in either the saturated regime or the damping regime of the curve of growth, not the linear regime.  Lines in the saturated or damping regime are especially susceptible to magnetic enhancement via Zeeman splitting: as saturated lines split, the equivalent width increases.  The location on the curve of growth is further evidence that the increase in equivalent width with Rossby number is due to magnetic enhancement and not a non-magnetic source, which we explore further in the next section.

It is curious that the spread in equivalent width values for a given line in Figure \ref{fig:cog} does not appear to depend on the effective \lande $g$ factor ($g_{\rm eff}$) of the corresponding transition.  The lines with the lowest $g_{\rm eff}$, \ti (5) and \ti (6), both under 1, show just as much of a spread in equivalent width as the transitions with \lande g factors greater than 1.  We cannot explain this apparent lack of dependence on $g_{\rm eff}$.  We note that the $g_{\rm eff}$ factors for these particular \ti lines are well known.  \ti lines with  similar lower and upper energy levels have been used in previous investigations of magnetic fields on M dwarfs \citep[][]{Shulyak2017,Shulyak2019} and \ti (2) has been used to measure magnetic fields on sunspots \citep[][]{Shulyak2010}.  We leave a detailed investigation into the magnetic field strength associated with the observed Zeeman enhancement to a future paper.

We note that the equivalent widths show no obvious correlation with stellar mass, and correspondingly, stellar effective temperature.  Being in the saturated regime of the curve of growth, the equivalent widths should be relatively insensitive to temperature, as the column density of \ti atoms in these particular electron configurations should have a weak impact on the measured equivalent width.  Indeed, in Figure \ref{fig:ew_v_mass}, we see a weak dependence on stellar mass.

\subsection{Non-magnetic Sources for the Trend}

We rule out the following sources for the trend in equivalent width versus Rossby number: rotational broadening, Ti abundance, and mass/temperature/gravity.  Regarding rotational broadening, objects were specifically chosen such that rotational broadening would not significantly affect the resulting line profiles.  We also argue that Ti abundance cannot be responsible either, although differences in Ti abundance may well be responsible for the scatter.  Ti abundance does not explain the saturation seen for $Ro \lesssim 0.1$, consistent with H$\alpha$ and X-ray emission.

Temperature, mass and surface gravity are other potential explanations for the trends and saturation; however, we include a wide range of stellar masses (and corresponding temperatures and surface gravities) in this sample.  In Figure \ref{fig:ew_v_mass}, we see similar behavior in equivalent width versus Rossby number as a function of mass.  We therefore rule this out as an explanation.

\subsection{Saturation Value}

We aim to compare the saturation behavior of \ti equivalent widths to chromospheric and coronal emission.  For each of the six lines, we fit a function to the equivalent width versus Rossby number using same function form used for H$\alpha$ and X-ray emission \citep[][]{Newton2017,Wright2011,Wright2018}:

\begin{equation}
EW = \begin{cases}
EW_{sat}, & Ro \leq Ro_{\rm sat} \\
C R_o^{\beta}, & R_o>R_{\rm sat} 
\end{cases}
\end{equation}

\noindent where $C$ is fixed to a constant to ensure continuity between regimes.  We note that in Figure \ref{fig:ew_v_ro}, the equivalent widths are on a linear scale; whereas chromospheric and coronal emission are often varying on a logarithmic scale, resulting in very different values for $\beta$.  For each line, we fit this functional form to the data using a Levenburg-Marquardt optimization routine \citep[][]{Levenberg1944,Marquardt1963}.  We removed the outliers discussed above from the fitted data.

\begin{figure*}[]
\includegraphics[width=0.99\textwidth]{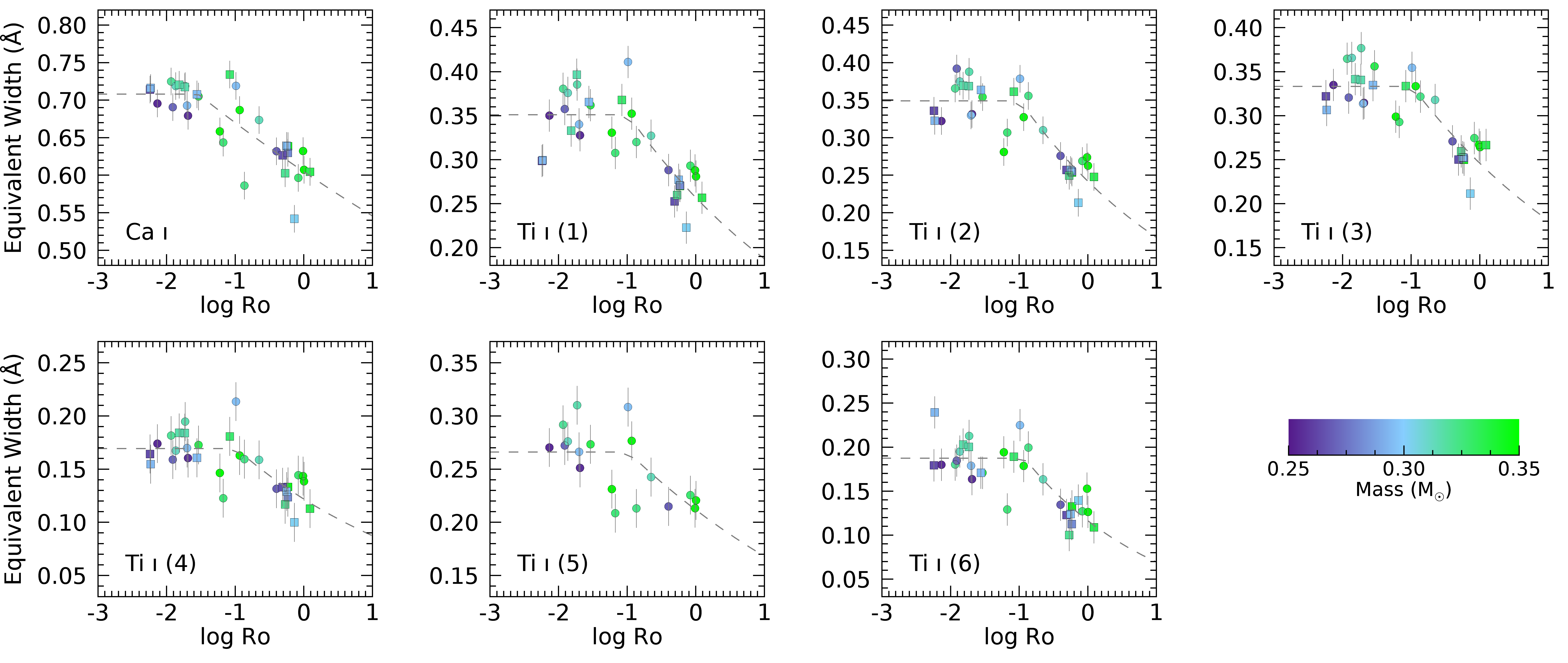}
\caption{Same as Figure \ref{fig:ew_v_ro}, but including model fits to the saturation with Rossby number (dashed lines).  We also limit the targets to those with masses between 0.25 and 0.35 $M_\sun$, where we have a wide range of Rossby numbers in our sample, to reduce the effect of stellar mass and tempearture.\label{fig:ew_v_ro_fit}}
\end{figure*}

In order to correct for the small variations in equivalent width with stellar mass, as seen in Figure \ref{fig:ew_v_mass}, we limited our targets to those with masses between 0.25 and 0.35 $M_\sun$ in the fitting routine. Figure \ref{fig:ew_v_ro_fit} shows the resulting fits and Table \ref{tab:fits} lists the resulting values.  We include a calculation of the difference in Bayesian Inference Criterion \citep[BIC,][]{Schwarz1978} between the fitted saturation model and a fitted log-linear model (i.e. a line versus $\log Ro$). The BIC parameter favors models with better fit the data while penalizing models for excessive model parameters.  Values lower than -10 indicate a strong preference for the saturation model, values between than -10 and 0 indicate weak-to-no preference, and values between 0 and 10 indicate weak-to-no preference for the log-linear model.  The saturation model is strongly preferred for \ti (1), (2), and (3).  For \ca and \ti (4) and (6) the saturation model is weakly preferred, and for \ti (5) a log-linear model is somewhat preferred.

We find a remarkably similar saturation value of $Ro$ for all of the \ti lines near $Ro=0.1$.  The saturation values are similar to what is seen in chromospheric saturation of M dwarf stars.  \citet{Newton2017} found a value of $Ro_{\rm sat}=0.21 \pm 0.02$ for saturation of H$\alpha$ with Rossby number, and \citet{Wright2018} found a value of $Ro_{\rm sat}=0.14^{+0.08}_{-0.04}$ for X-ray emission for fully convective M dwarfs.  \citet{Douglas2014} found a value of 0.11$^{+0.02}){-0.03}$ for K and M dwarfs in the Hyades cluster.  The similarity in saturation value for the \ti equivalent widths and H$\alpha$ and X-ray emission provides strong evidence that the \ti lines are magnetically enhanced.

\begin{table}[]
    \centering
    \begin{tabular}{lcccr}
    \hline 
    Line & $C$ & $\beta$ & $R_{\rm sat}$ & $\Delta$BIC \\
    \hline 
\ion{Ca}{1} &  0.7079 & -0.0445 &  0.0288 &    -1.6 \\
\ion{Ti}{1} (1) &  0.3510 & -0.1370 &  0.1024 &   -35.5 \\
\ion{Ti}{1} (2) &  0.3489 & -0.1608 &  0.1027 &   -21.8 \\
\ion{Ti}{1} (3) &  0.3332 & -0.1336 &  0.1027 &   -13.8 \\
\ion{Ti}{1} (4) &  0.1693 & -0.1446 &  0.1027 &    -2.7 \\
\ion{Ti}{1} (5) &  0.2660 & -0.1003 &  0.1022 &     6.9 \\
\ion{Ti}{1} (6) &  0.1875 & -0.2236 &  0.1182 &    -4.8 \\

\hline
\end{tabular}
    \caption{Fitted parameters.}
    \label{tab:fits}
\end{table}

\section{Discussion}\label{sec:discussion}

In the context of existing measurements of magnetic fields for large samples of cool stars \citep[e.g.][]{Reiners2009,Shulyak2019,Afram2019}, we argue the impact of this work is the empirical detection of saturation of a photospheric signature with Rossby number.  Previous efforts in this area have involved complex modeling of spectra to determine magnetic field strengths; however, the conversion from spectra to magnetic field is not straightforward, and the authors are unaware of a clear detection of saturation in equivalent width in absorption with Rossby number for M dwarfs in the literature.  As for the physical source of the Zeeman enhancement, it is unclear whether it arises from isolated magnetic spots, a global magnetic field, or a combination of the two.

In combination with the results on FeH lines from \citet[][]{Reiners2009}, these results strongly suggest that the saturation mechanism occurs at or below the stellar photosphere and not the chromosphere or corona.  That is to say the magnetic fields in the photosphere itself are saturated.  However, the fundamental nature of that mechanism remains a mystery.  The saturation may well be the fundamental conversion of rotational/convective sheering into a magnetic field in the first place.  

\subsection{Metallicity}

Recently, \citet[]{Passegger2019} made the important point that using magnetically sensitive lines may corrupt metallicity determinations unless Zeeman enhancement is modelled. In their work, they determined metallicities of M dwarfs in the CARMENES data using model atmospheres and magnetically insensitive lines. \citet{Shulyak2019} made a similar point, stating that measurements of magnetic field strength using Zeeman enhancement require an estimate of the metallicity.  In this work, we side step the issue of measurement metallicity or magnetic fields, instead focusing on the observational evidence for a saturation mechanism at or below the M dwarf photosphere.  However, metallicity may add to the scatter in Figure \ref{fig:ew_v_ro}.

Nevertheless, it is important to discuss the implications this work has for metallicity determinations.  \citet{Veyette2017} and \citet{Veyette2018} used the same \ti lines to determine the $\alpha$ enrichment of M dwarf stars and measure chemical-kinematic ages of exoplanet hosts without considering magnetic effects.  However, \citet{Veyette2017} found that M dwarfs known to have enhanced Ti (based on the Ti content of FGK companions) did in fact show larger Ti equivalent widths and followed the equivalent width-metallicity trends expected from non-magnetic stellar atmosphere models (see their Figure 4).  That is to say: without considering magnetic effects, \citet{Veyette2017} found that equivalent widths of \ti lines trace Ti abundance.  In this work, without considering metallicity effects, we find that equivalent widths of \ti lines trace magnetic enhancement.  Combining these results, it appears that magnetic enhancement is a source of scatter in the metallicity relations, and vice versa: metallicity is a source of scatter in magnetic enhancement.  This is entirely empirical: we see these trends without employing any M dwarf atmospheric models, which are subject to their own peculiar systematic errors.

\subsection{Radius Inflation}

This work is part of a larger paper series on the nature of discrepancies between modelled and measured radii of M dwarf stars and what role, if any, magnetic fields play in that discrepancy \citep[``Magnetic Inflation and Stellar Mass'', ][]{Han2017,Kesseli2018,Healy2019,Han2019}.  A prominent theory for the discrepancies involves magnetic inhibition of convection in the outer, super-adiabatic layer of the star \citep[][]{Chabrier2007}.  If the magnetic fields in the photosphere saturate, as implied by our results, it suggests that M dwarfs of similar masses in the saturated regime ($Ro \lesssim 0.1$) should have the same degree of radius inflation with respect to a fiducial value: either predictions from non-magnetic evolutionary models or non-active M dwarfs with high Rossby number.

\subsection{Magnetic Variability}

We note that magnetic activity indicators are known to change in time as stars rotate and go through their activity cycles.  Time-domain observations of activity indicators (including the equivalent widths of magnetically sensitive lines like those analyzed here), will enable a global and cycle-integrated view of M dwarf surface fields, rather than a snapshot in time.  With long-term, high-cadence magnetic spectroscopic monitoring of M dwarf stars, similar to the Mt. Wilson program to monitor the \ion{Ca}{2} H and K lines in nearby sun-like stars \citep[][]{Wilson1966}, we can further test predictions concerning the role of magnetic activity of M dwarf radii.

\acknowledgments

We would like to thank the anonymous referee for providing helpful comments on the manuscript.  P.S.M. would like to thank Svetlana Berdyugina, Denis Shulyak, Matthew Browning and Gregory Feiden for helpful discussions regarding this work.  P.S.M acknowledges support for this work from the National Science Foundation under Grant No. 1716260.  Additional support was provided by a NASA Keck PI Data Award, administered by the NASA Exoplanet Science Institute (NExSci).  NExScI is sponsored by NASA's Exoplanet Program and operated by the California Institute of Technology in coordination with the Jet Propulsion Laboratory.  C.T. acknowledges support from NASA through the NASA Hubble Fellowship grant HST-HF2-51447.001-A  awarded by the Space Telescope Science Institute, which is operated by the Association of Universities for Research in Astronomy, Inc., for NASA, under contract NAS5-26555.

This research benefited from discussions during the \textit{Better Stars, Better Planets: Exploiting the Stellar-Exoplanetary Synergy} program held at the Kavli Institute for Theoretical Physics (KITP) in Santa Barbara, CA.  P.S.M and E.N. thank the program coordinators for organizing such a  successful program.  KITP is supported in part by the National Science Foundation under Grant No. NSF PHY-1748958.

The \texttt{NSDRP} software was developed by the Keck Observatory Archive (KOA), which is operated by the W. M. Keck Observatory and NExScI, under contract with the National Aeronautics and Space Administration.  This research made use of the Massachusetts Green High Performance Computing Center in Holyoke, MA.  This work has made use of the VALD database, operated at Uppsala University, the Institute of Astronomy RAS in Moscow, and the University of Vienna.  This research has made use of the SIMBAD database, operated at CDS, Strasbourg, France.

Some of the data presented herein were obtained at the W.M. Keck Observatory, which is operated as a scientific partnership among the California Institute of Technology, the University of California and the National Aeronautics and Space Administration. The Observatory was made possible by the generous financial support of the W.M. Keck Foundation.  We would like to personally thank Emily Martin and Gregory Doppmann for their assistance with operating NIRSPEC during the observing run.

The authors wish to recognize and acknowledge the very significant cultural role and reverence that the summit of Mauna Kea has always had within the indigenous Hawaiian community.  We are most fortunate to have the opportunity to conduct observations from this mountain.

\vspace{5mm}
\facilities{Keck:II (NIRSPEC), CAO:3.5m (CARMENES)}

\software{\texttt{analyze\_NIRSPEC1} \citep{Veyette2017},
\texttt{astropy} \citep{Astropy2013},  
\texttt{NSDRP} \citep{Tran2016}}

\bibliography{references}

\begin{thebibliography}{}
\expandafter\ifx\csname natexlab\endcsname\relax\def\natexlab#1{#1}\fi
\providecommand{\url}[1]{\href{#1}{#1}}
\providecommand{\dodoi}[1]{doi:~\href{http://doi.org/#1}{\nolinkurl{#1}}}
\providecommand{\doeprint}[1]{\href{http://ascl.net/#1}{\nolinkurl{http://ascl.net/#1}}}
\providecommand{\doarXiv}[1]{\href{https://arxiv.org/abs/#1}{\nolinkurl{https://arxiv.org/abs/#1}}}

\bibitem[{{Afram} \& {Berdyugina}(2019)}]{Afram2019}
{Afram}, N., \& {Berdyugina}, S.~V. 2019, arXiv e-prints, arXiv:1908.00076.
\newblock \doarXiv{1908.00076}

\bibitem[{{Allard} {et~al.}(2012){Allard}, {Homeier}, \&
  {Freytag}}]{Allard2012}
{Allard}, F., {Homeier}, D., \& {Freytag}, B. 2012, Philosophical Transactions
  of the Royal Society of London Series A, 370, 2765,
  \dodoi{10.1098/rsta.2011.0269}

\bibitem[{{Astropy Collaboration} {et~al.}(2013){Astropy Collaboration},
  {Robitaille}, {Tollerud}, {Greenfield}, {Droettboom}, {Bray}, {Aldcroft},
  {Davis}, {Ginsburg}, {Price-Whelan}, {Kerzendorf}, {Conley}, {Crighton},
  {Barbary}, {Muna}, {Ferguson}, {Grollier}, {Parikh}, {Nair}, {Unther},
  {Deil}, {Woillez}, {Conseil}, {Kramer}, {Turner}, {Singer}, {Fox}, {Weaver},
  {Zabalza}, {Edwards}, {Azalee Bostroem}, {Burke}, {Casey}, {Crawford},
  {Dencheva}, {Ely}, {Jenness}, {Labrie}, {Lim}, {Pierfederici}, {Pontzen},
  {Ptak}, {Refsdal}, {Servillat}, \& {Streicher}}]{Astropy2013}
{Astropy Collaboration}, {Robitaille}, T.~P., {Tollerud}, E.~J., {et~al.} 2013,
  \aap, 558, A33, \dodoi{10.1051/0004-6361/201322068}

\bibitem[{{Basri} {et~al.}(1992){Basri}, {Marcy}, \& {Valenti}}]{Basri1992}
{Basri}, G., {Marcy}, G.~W., \& {Valenti}, J.~A. 1992, \apj, 390, 622,
  \dodoi{10.1086/171312}

\bibitem[{Berger(2006)}]{Berger2006}
Berger, E. 2006, The Astrophysical Journal, 648, 629, \dodoi{10.1086/505787}

\bibitem[{{Bouvier} {et~al.}(2014){Bouvier}, {Matt}, {Mohanty}, {Scholz},
  {Stassun}, \& {Zanni}}]{Bouvier2014}
{Bouvier}, J., {Matt}, S.~P., {Mohanty}, S., {et~al.} 2014, in Protostars and
  Planets VI, ed. H.~{Beuther}, R.~S. {Klessen}, C.~P. {Dullemond}, \&
  T.~{Henning}, 433

\bibitem[{{Chabrier} {et~al.}(2007){Chabrier}, {Gallardo}, \&
  {Baraffe}}]{Chabrier2007}
{Chabrier}, G., {Gallardo}, J., \& {Baraffe}, I. 2007, \aap, 472, L17,
  \dodoi{10.1051/0004-6361:20077702}

\bibitem[{{Charbonneau} {et~al.}(2009){Charbonneau}, {Berta}, {Irwin}, {Burke},
  {Nutzman}, {Buchhave}, {Lovis}, {Bonfils}, {Latham}, \&
  {Udry}}]{Charbonneau2009}
{Charbonneau}, D., {Berta}, Z.~K., {Irwin}, J., {et~al.} 2009, \nat, 462, 891,
  \dodoi{10.1038/nature08679}

\bibitem[{{Cushing} {et~al.}(2004){Cushing}, {Vacca}, \&
  {Rayner}}]{Cushing2004}
{Cushing}, M.~C., {Vacca}, W.~D., \& {Rayner}, J.~T. 2004, \pasp, 116, 362,
  \dodoi{10.1086/382907}

\bibitem[{{Cutri} {et~al.}(2003){Cutri}, {Skrutskie}, {van Dyk}, {Beichman},
  {Carpenter}, {Chester}, {Cambresy}, {Evans}, {Fowler}, {Gizis}, {Howard},
  {Huchra}, {Jarrett}, {Kopan}, {Kirkpatrick}, {Light}, {Marsh}, {McCallon},
  {Schneider}, {Stiening}, {Sykes}, {Weinberg}, {Wheaton}, {Wheelock}, \&
  {Zacarias}}]{Cutri2003}
{Cutri}, R.~M., {Skrutskie}, M.~F., {van Dyk}, S., {et~al.} 2003, {2MASS All
  Sky Catalog of point sources.}

\bibitem[{{D{\'\i}ez Alonso} {et~al.}(2019){D{\'\i}ez Alonso}, {Caballero},
  {Montes}, {de Cos Juez}, {Dreizler}, {Dubois}, {Jeffers}, {Lalitha}, {Naves},
  {Reiners}, {Ribas}, {Vanaverbeke}, {Amado}, {B{\'e}jar},
  {Cort{\'e}s-Contreras}, {Herrero}, {Hidalgo}, {K{\"u}rster}, {Logie},
  {Quirrenbach}, {Rau}, {Seifert}, {Sch{\"o}fer}, \& {Tal-Or}}]{DiezAlonso2019}
{D{\'\i}ez Alonso}, E., {Caballero}, J.~A., {Montes}, D., {et~al.} 2019, \aap,
  621, A126, \dodoi{10.1051/0004-6361/201833316}

\bibitem[{{Dittmann} {et~al.}(2016){Dittmann}, {Irwin}, {Charbonneau}, \&
  {Newton}}]{Dittmann2016}
{Dittmann}, J.~A., {Irwin}, J.~M., {Charbonneau}, D., \& {Newton}, E.~R. 2016,
  \apj, 818, 153, \dodoi{10.3847/0004-637X/818/2/153}

\bibitem[{{Dittmann} {et~al.}(2017){Dittmann}, {Irwin}, {Charbonneau},
  {Berta-Thompson}, {Newton}, {Latham}, {Latham}, {Esquerdo}, {Berlind}, \&
  {Calkins}}]{Dittmann2017}
{Dittmann}, J.~A., {Irwin}, J.~M., {Charbonneau}, D., {et~al.} 2017, \apj, 836,
  124, \dodoi{10.3847/1538-4357/836/1/124}

\bibitem[{{Douglas} {et~al.}(2014){Douglas}, {Ag{\"u}eros}, {Covey}, {Bowsher},
  {Bochanski}, {Cargile}, {Kraus}, {Law}, {Lemonias}, {Arce}, {Fierroz}, \&
  {Kundert}}]{Douglas2014}
{Douglas}, S.~T., {Ag{\"u}eros}, M.~A., {Covey}, K.~R., {et~al.} 2014, \apj,
  795, 161, \dodoi{10.1088/0004-637X/795/2/161}

\bibitem[{{Fowler} \& {Gatley}(1990)}]{Fowler1990}
{Fowler}, A.~M., \& {Gatley}, I. 1990, \apj, 353, L33, \dodoi{10.1086/185701}

\bibitem[{{France} {et~al.}(2016){France}, {Loyd}, {Youngblood}, {Brown},
  {Schneider}, {Hawley}, {Froning}, {Linsky}, {Roberge}, \&
  {Buccino}}]{France2016}
{France}, K., {Loyd}, R.~O.~P., {Youngblood}, A., {et~al.} 2016, \apj, 820, 89,
  \dodoi{10.3847/0004-637X/820/2/89}

\bibitem[{{Gaia Collaboration}(2018)}]{Gaia2018}
{Gaia Collaboration}. 2018, VizieR Online Data Catalog, I/345

\bibitem[{{Han} {et~al.}(2019){Han}, {Muirhead}, \& {Swift}}]{Han2019}
{Han}, E., {Muirhead}, P.~S., \& {Swift}, J.~J. 2019, \aj, 158, 111,
  \dodoi{10.3847/1538-3881/ab2ed7}

\bibitem[{{Han} {et~al.}(2017){Han}, {Muirhead}, {Swift}, {Baranec}, {Law},
  {Riddle}, {Atkinson}, {Mace}, \& {DeFelippis}}]{Han2017}
{Han}, E., {Muirhead}, P.~S., {Swift}, J.~J., {et~al.} 2017, \aj, 154, 100,
  \dodoi{10.3847/1538-3881/aa803c}

\bibitem[{{Healy} {et~al.}(2019){Healy}, {Han}, {Muirhead}, {Skiff}, {Polakis},
  {Rilinger}, \& {Swift}}]{Healy2019}
{Healy}, B.~F., {Han}, E., {Muirhead}, P.~S., {et~al.} 2019, \aj, 158, 89,
  \dodoi{10.3847/1538-3881/ab2fe5}

\bibitem[{{Horne}(1986)}]{Horne1986}
{Horne}, K. 1986, \pasp, 98, 609, \dodoi{10.1086/131801}

\bibitem[{{Jardine} \& {Unruh}(1999)}]{Jardine1999}
{Jardine}, M., \& {Unruh}, Y.~C. 1999, \aap, 346, 883

\bibitem[{{Kay} {et~al.}(2016){Kay}, {Opher}, \& {Kornbleuth}}]{Kay2016}
{Kay}, C., {Opher}, M., \& {Kornbleuth}, M. 2016, \apj, 826, 195,
  \dodoi{10.3847/0004-637X/826/2/195}

\bibitem[{{Kesseli} {et~al.}(2018){Kesseli}, {Muirhead}, {Mann}, \&
  {Mace}}]{Kesseli2018}
{Kesseli}, A.~Y., {Muirhead}, P.~S., {Mann}, A.~W., \& {Mace}, G. 2018, \aj,
  155, 225, \dodoi{10.3847/1538-3881/aabccb}

\bibitem[{{Kurucz}(2010)}]{K10}
{Kurucz}, R.~L. 2010, Robert L. Kurucz on-line database of observed and
  predicted atomic transitions

\bibitem[{Levenberg(1944)}]{Levenberg1944}
Levenberg, K. 1944, Quarterly of Applied Mathematics, 2, 164

\bibitem[{{Mann} {et~al.}(2015){Mann}, {Feiden}, {Gaidos}, {Boyajian}, \& {von
  Braun}}]{Mann2015}
{Mann}, A.~W., {Feiden}, G.~A., {Gaidos}, E., {Boyajian}, T., \& {von Braun},
  K. 2015, \apj, 804, 64, \dodoi{10.1088/0004-637X/804/1/64}

\bibitem[{{Mann} {et~al.}(2019){Mann}, {Dupuy}, {Kraus}, {Gaidos}, {Ansdell},
  {Ireland}, {Rizzuto}, {Hung}, {Dittmann}, \& {Factor}}]{Mann2019}
{Mann}, A.~W., {Dupuy}, T., {Kraus}, A.~L., {et~al.} 2019, \apj, 871, 63,
  \dodoi{10.3847/1538-4357/aaf3bc}

\bibitem[{Marquardt(1963)}]{Marquardt1963}
Marquardt, D.~W. 1963, SIAM Journal of Applied Mathematics, 11, 431

\bibitem[{{Martin} {et~al.}(2018){Martin}, {Fitzgerald}, {McLean}, {Doppmann},
  {Kassis}, {Aliado}, {Canfield}, {Johnson}, {Kress}, \&
  {Lanclos}}]{Martin2018}
{Martin}, E.~C., {Fitzgerald}, M.~P., {McLean}, I.~S., {et~al.} 2018, in
  Society of Photo-Optical Instrumentation Engineers (SPIE) Conference Series,
  Vol. 10702, Ground-based and Airborne Instrumentation for Astronomy VII,
  107020A

\bibitem[{{McLean} {et~al.}(1998){McLean}, {Becklin}, {Bendiksen}, {Brims},
  {Canfield}, {Figer}, {Graham}, {Hare}, {Lacayanga}, {Larkin}, {Larson},
  {Levenson}, {Magnone}, {Teplitz}, \& {Wong}}]{McLean1998}
{McLean}, I.~S., {Becklin}, E.~E., {Bendiksen}, O., {et~al.} 1998, in
  \procspie, Vol. 3354, Infrared Astronomical Instrumentation, ed. A.~M.
  {Fowler}, 566--578

\bibitem[{{Muirhead} {et~al.}(2013){Muirhead}, {Vanderburg}, {Shporer},
  {Becker}, {Swift}, {Lloyd}, {Fuller}, {Zhao}, {Hinkley}, {Pineda}, {Bottom},
  {Howard}, {von Braun}, {Boyajian}, {Law}, {Baranec}, {Riddle}, {Ramaprakash},
  {Tendulkar}, {Bui}, {Burse}, {Chordia}, {Das}, {Dekany}, {Punnadi}, \&
  {Johnson}}]{Muirhead2013}
{Muirhead}, P.~S., {Vanderburg}, A., {Shporer}, A., {et~al.} 2013, \apj, 767,
  111, \dodoi{10.1088/0004-637X/767/2/111}

\bibitem[{{Newton} {et~al.}(2015){Newton}, {Charbonneau}, {Irwin}, \&
  {Mann}}]{Newton2015}
{Newton}, E.~R., {Charbonneau}, D., {Irwin}, J., \& {Mann}, A.~W. 2015, \apj,
  800, 85, \dodoi{10.1088/0004-637X/800/2/85}

\bibitem[{{Newton} {et~al.}(2017){Newton}, {Irwin}, {Charbonneau}, {Berlind},
  {Calkins}, \& {Mink}}]{Newton2017}
{Newton}, E.~R., {Irwin}, J., {Charbonneau}, D., {et~al.} 2017, \apj, 834, 85,
  \dodoi{10.3847/1538-4357/834/1/85}

\bibitem[{{Noyes} {et~al.}(1984){Noyes}, {Hartmann}, {Baliunas}, {Duncan}, \&
  {Vaughan}}]{Noyes1984}
{Noyes}, R.~W., {Hartmann}, L.~W., {Baliunas}, S.~L., {Duncan}, D.~K., \&
  {Vaughan}, A.~H. 1984, \apj, 279, 763, \dodoi{10.1086/161945}

\bibitem[{{Pallavicini} {et~al.}(1981){Pallavicini}, {Golub}, {Rosner},
  {Vaiana}, {Ayres}, \& {Linsky}}]{Pallavicini1981}
{Pallavicini}, R., {Golub}, L., {Rosner}, R., {et~al.} 1981, The Astrophysical
  Journal, 248, 279, \dodoi{10.1086/159152}

\bibitem[{{Passegger} {et~al.}(2019){Passegger}, {Schweitzer}, {Shulyak},
  {Nagel}, {Hauschildt}, {Reiners}, {Amado}, {Caballero},
  {Cort{\'e}s-Contreras}, {Dom{\'\i}nguez-Fern{\'a}ndez}, {Quirrenbach},
  {Ribas}, {Azzaro}, {Anglada-Escud{\'e}}, {Bauer}, {B{\'e}jar}, {Dreizler},
  {Guenther}, {Henning}, {Jeffers}, {Kaminski}, {K{\"u}rster}, {Lafarga},
  {Mart{\'\i}n}, {Montes}, {Morales}, {Schmitt}, \&
  {Zechmeister}}]{Passegger2019}
{Passegger}, V.~M., {Schweitzer}, A., {Shulyak}, D., {et~al.} 2019, Astronomy
  and Astrophysics, 627, A161, \dodoi{10.1051/0004-6361/201935679}

\bibitem[{{Paulson} {et~al.}(2006){Paulson}, {Allred}, {Anderson}, {Hawley},
  {Cochran}, \& {Yelda}}]{Paulson2006}
{Paulson}, D.~B., {Allred}, J.~C., {Anderson}, R.~B., {et~al.} 2006, \pasp,
  118, 227, \dodoi{10.1086/499497}

\bibitem[{{Piskunov} {et~al.}(1995){Piskunov}, {Kupka}, {Ryabchikova}, {Weiss},
  \& {Jeffery}}]{Piskunov1995}
{Piskunov}, N.~E., {Kupka}, F., {Ryabchikova}, T.~A., {Weiss}, W.~W., \&
  {Jeffery}, C.~S. 1995, \aaps, 112, 525

\bibitem[{{Pizzolato} {et~al.}(2003){Pizzolato}, {Maggio}, {Micela},
  {Sciortino}, \& {Ventura}}]{Pizzolato2003}
{Pizzolato}, N., {Maggio}, A., {Micela}, G., {Sciortino}, S., \& {Ventura}, P.
  2003, \aap, 397, 147, \dodoi{10.1051/0004-6361:20021560}

\bibitem[{{Quirrenbach} {et~al.}(2016){Quirrenbach}, {Amado}, {Caballero},
  {Mundt}, {Reiners}, {Ribas}, {Seifert}, {Abril}, {Aceituno},
  {Alonso-Floriano}, {Anwand-Heerwart}, {Azzaro}, {Bauer}, {Barrado},
  {Becerril}, {Bejar}, {Benitez}, {Berdinas}, {Brinkm{\"o}ller}, {Cardenas},
  {Casal}, {Claret}, {Colom{\'e}}, {Cortes-Contreras}, {Czesla}, {Doellinger},
  {Dreizler}, {Feiz}, {Fernandez}, {Ferro}, {Fuhrmeister}, {Galadi},
  {Gallardo}, {G{\'a}lvez-Ortiz}, {Garcia-Piquer}, {Garrido}, {Gesa},
  {G{\'o}mez Galera}, {Gonz{\'a}lez Hern{\'a}ndez}, {Gonzalez Peinado},
  {Gr{\"o}zinger}, {Gu{\`a}rdia}, {Guenther}, {de Guindos}, {Hagen}, {Hatzes},
  {Hauschildt}, {Helmling}, {Henning}, {Hermann}, {Hern{\'a}ndez Arabi},
  {Hern{\'a}ndez Casta{\~n}o}, {Hern{\'a}ndez Hernando}, {Herrero}, {Huber},
  {Huber}, {Huke}, {Jeffers}, {de Juan}, {Kaminski}, {Kehr}, {Kim}, {Klein},
  {Kl{\"u}ter}, {K{\"u}rster}, {Lafarga}, {Lara}, {Lamert}, {Laun},
  {Launhardt}, {Lemke}, {Lenzen}, {Llamas}, {Lopez del Fresno},
  {L{\'o}pez-Puertas}, {L{\'o}pez-Santiago}, {Lopez Salas}, {Magan
  Madinabeitia}, {Mall}, {Mandel}, {Mancini}, {Marin Molina}, {Maroto
  Fern{\'a}ndez}, {Mart{\'{\i}}n}, {Mart{\'{\i}}n-Ruiz}, {Marvin}, {Mathar},
  {Mirabet}, {Montes}, {Morales}, {Morales Mu{\~n}oz}, {Nagel}, {Naranjo},
  {Nowak}, {Palle}, {Panduro}, {Passegger}, {Pavlov}, {Pedraz}, {Perez},
  {P{\'e}rez-Medialdea}, {Perger}, {Pluto}, {Ram{\'o}n}, {Rebolo}, {Redondo},
  {Reffert}, {Reinhart}, {Rhode}, {Rix}, {Rodler}, {Rodr{\'{\i}}guez},
  {Rodr{\'{\i}}guez L{\'o}pez}, {Rohloff}, {Rosich}, {Sanchez Carrasco},
  {Sanz-Forcada}, {Sarkis}, {Sarmiento}, {Sch{\"a}fer}, {Schiller}, {Schmidt},
  {Schmitt}, {Sch{\"o}fer}, {Schweitzer}, {Shulyak}, {Solano}, {Stahl},
  {Storz}, {Tabernero}, {Tala}, {Tal-Or}, {Ulbrich}, {Veredas}, {Vico Linares},
  {Vilardell}, {Wagner}, {Winkler}, {Zapatero Osorio}, {Zechmeister},
  {Ammler-von Eiff}, {Anglada-Escud{\'e}}, {del Burgo}, {Garcia-Vargas},
  {Klutsch}, {Lizon}, {Lopez-Morales}, {Ofir}, {P{\'e}rez-Calpena}, {Perryman},
  {S{\'a}nchez-Blanco}, {Strachan}, {St{\"u}rmer}, {Su{\'a}rez}, {Trifonov},
  {Tulloch}, \& {Xu}}]{Quirrenbach2016}
{Quirrenbach}, A., {Amado}, P.~J., {Caballero}, J.~A., {et~al.} 2016, in
  \procspie, Vol. 9908, Ground-based and Airborne Instrumentation for Astronomy
  VI, 990812

\bibitem[{{Reiners} {et~al.}(2009){Reiners}, {Basri}, \&
  {Browning}}]{Reiners2009}
{Reiners}, A., {Basri}, G., \& {Browning}, M. 2009, \apj, 692, 538,
  \dodoi{10.1088/0004-637X/692/1/538}

\bibitem[{{Reiners} {et~al.}(2018){Reiners}, {Zechmeister}, {Caballero},
  {Ribas}, {Morales}, {Jeffers}, {Sch{\"o}fer}, {Tal-Or}, {Quirrenbach},
  {Amado}, {Kaminski}, {Seifert}, {Abril}, {Aceituno}, {Alonso-Floriano},
  {Ammler-von Eiff}, {Antona}, {Anglada-Escud{\'e}}, {Anwand-Heerwart},
  {Arroyo-Torres}, {Azzaro}, {Baroch}, {Barrado}, {Bauer}, {Becerril},
  {B{\'e}jar}, {Ben{\'\i}tez}, {Berdinas̃}, {Bergond}, {Bl{\"u}mcke},
  {Brinkm{\"o}ller}, {del Burgo}, {Cano}, {C{\'a}rdenas V{\'a}zquez}, {Casal},
  {Cifuentes}, {Claret}, {Colom{\'e}}, {Cort{\'e}s-Contreras}, {Czesla},
  {D{\'\i}ez-Alonso}, {Dreizler}, {Feiz}, {Fern{\'a}ndez}, {Ferro},
  {Fuhrmeister}, {Galad{\'\i}-Enr{\'\i}quez}, {Garcia-Piquer}, {Garc{\'\i}a
  Vargas}, {Gesa}, {G{\'o}mez Galera}, {Gonz{\'a}lez Hern{\'a}ndez},
  {Gonz{\'a}lez-Peinado}, {Gr{\"o}zinger}, {Grohnert}, {Gu{\`a}rdia},
  {Guenther}, {Guijarro}, {de Guindos}, {Guti{\'e}rrez-Soto}, {Hagen},
  {Hatzes}, {Hauschildt}, {Hedrosa}, {Helmling}, {Henning}, {Hermelo},
  {Hern{\'a}ndez Arab{\'\i}}, {Hern{\'a}ndez Casta{\~n}o}, {Hern{\'a}ndez
  Hernando}, {Herrero}, {Huber}, {Huke}, {Johnson}, {de Juan}, {Kim}, {Klein},
  {Kl{\"u}ter}, {Klutsch}, {K{\"u}rster}, {Lafarga}, {Lamert}, {Lamp{\'o}n},
  {Lara}, {Laun}, {Lemke}, {Lenzen}, {Launhardt}, {L{\'o}pez del Fresno},
  {L{\'o}pez-Gonz{\'a}lez}, {L{\'o}pez-Puertas}, {L{\'o}pez Salas},
  {L{\'o}pez-Santiago}, {Luque}, {Mag{\'a}n Madinabeitia}, {Mall}, {Mancini},
  {Mand el}, {Marfil}, {Mar{\'\i}n Molina}, {Maroto Fern{\'a}ndez},
  {Mart{\'\i}n}, {Mart{\'\i}n-Ruiz}, {Marvin}, {Mathar}, {Mirabet}, {Montes},
  {Moreno-Raya}, {Moya}, {Mundt}, {Nagel}, {Naranjo}, {Nortmann}, {Nowak},
  {Ofir}, {Oreiro}, {Pall{\'e}}, {Pand uro}, {Pascual}, {Passegger}, {Pavlov},
  {Pedraz}, {P{\'e}rez-Calpena}, {P{\'e}rez Medialdea}, {Perger}, {Perryman},
  {Pluto}, {Rabaza}, {Ram{\'o}n}, {Rebolo}, {Redondo}, {Reffert}, {Reinhart},
  {Rhode}, {Rix}, {Rodler}, {Rodr{\'\i}guez}, {Rodr{\'\i}guez-L{\'o}pez},
  {Rodr{\'\i}guez Trinidad}, {Rohloff}, {Rosich}, {Sadegi},
  {S{\'a}nchez-Blanco}, {S{\'a}nchez Carrasco}, {S{\'a}nchez-L{\'o}pez},
  {Sanz-Forcada}, {Sarkis}, {Sarmiento}, {Sch{\"a}fer}, {Schmitt}, {Schiller},
  {Schweitzer}, {Solano}, {Stahl}, {Strachan}, {St{\"u}rmer}, {Su{\'a}rez},
  {Tabernero}, {Tala}, {Trifonov}, {Tulloch}, {Ulbrich}, {Veredas}, {Vico
  Linares}, {Vilardell}, {Wagner}, {Winkler}, {Wolthoff}, {Xu}, {Yan}, \&
  {Zapatero Osorio}}]{Reiners2018}
{Reiners}, A., {Zechmeister}, M., {Caballero}, J.~A., {et~al.} 2018, \aap, 612,
  A49, \dodoi{10.1051/0004-6361/201732054}

\bibitem[{{Ryabchikova} {et~al.}(2015){Ryabchikova}, {Piskunov}, {Kurucz},
  {Stempels}, {Heiter}, {Pakhomov}, \& {Barklem}}]{Ryabchikova2015}
{Ryabchikova}, T., {Piskunov}, N., {Kurucz}, R.~L., {et~al.} 2015, \physscr,
  90, 054005, \dodoi{10.1088/0031-8949/90/5/054005}

\bibitem[{{Savitzky} \& {Golay}(1964)}]{Savitzky1964}
{Savitzky}, A., \& {Golay}, M.~J.~E. 1964, Analytical Chemistry, 36, 1627

\bibitem[{{Schwarz}(1978)}]{Schwarz1978}
{Schwarz}, G. 1978, Annals of Statistics, 6, 461

\bibitem[{{Shulyak} {et~al.}(2017){Shulyak}, {Reiners}, {Engeln}, {Malo},
  {Yadav}, {Morin}, \& {Kochukhov}}]{Shulyak2017}
{Shulyak}, D., {Reiners}, A., {Engeln}, A., {et~al.} 2017, Nature Astronomy, 1,
  0184, \dodoi{10.1038/s41550-017-0184}

\bibitem[{{Shulyak} {et~al.}(2010){Shulyak}, {Reiners}, {Wende}, {Kochukhov},
  {Piskunov}, \& {Seifahrt}}]{Shulyak2010}
{Shulyak}, D., {Reiners}, A., {Wende}, S., {et~al.} 2010, \aap, 523, A37,
  \dodoi{10.1051/0004-6361/201015229}

\bibitem[{{Shulyak} {et~al.}(2019){Shulyak}, {Reiners}, {Nagel}, {Tal-Or},
  {Caballero}, {Zechmeister}, {B{\'e}jar}, {Cort{\'e}s-Contreras}, {Martin}, \&
  {Kaminski}}]{Shulyak2019}
{Shulyak}, D., {Reiners}, A., {Nagel}, E., {et~al.} 2019, arXiv e-prints,
  arXiv:1904.12762.
\newblock \doarXiv{1904.12762}

\bibitem[{{Skinner}(2015)}]{Skinner2015}
{Skinner}, J.~N. 2015, PhD thesis, Dartmouth College

\bibitem[{{Skrutskie} {et~al.}(2006){Skrutskie}, {Cutri}, {Stiening},
  {Weinberg}, {Schneider}, {Carpenter}, {Beichman}, {Capps}, {Chester},
  {Elias}, {Huchra}, {Liebert}, {Lonsdale}, {Monet}, {Price}, {Seitzer},
  {Jarrett}, {Kirkpatrick}, {Gizis}, {Howard}, {Evans}, {Fowler}, {Fullmer},
  {Hurt}, {Light}, {Kopan}, {Marsh}, {McCallon}, {Tam}, {Van Dyk}, \&
  {Wheelock}}]{Skrutskie2006}
{Skrutskie}, M.~F., {Cutri}, R.~M., {Stiening}, R., {et~al.} 2006, \aj, 131,
  1163, \dodoi{10.1086/498708}

\bibitem[{{Soderblom} {et~al.}(1993){Soderblom}, {Stauffer}, {Hudon}, \&
  {Jones}}]{Soderblom1993}
{Soderblom}, D.~R., {Stauffer}, J.~R., {Hudon}, J.~D., \& {Jones}, B.~F. 1993,
  \apjs, 85, 315, \dodoi{10.1086/191767}

\bibitem[{{Stelzer} {et~al.}(2016){Stelzer}, {Damasso}, {Scholz}, \&
  {Matt}}]{Stelzer2016}
{Stelzer}, B., {Damasso}, M., {Scholz}, A., \& {Matt}, S.~P. 2016, \mnras, 463,
  1844, \dodoi{10.1093/mnras/stw1936}

\bibitem[{{Stewart} {et~al.}(1988){Stewart}, {Innis}, {Slee}, {Nelson}, \&
  {Wright}}]{Stewart1988}
{Stewart}, R.~T., {Innis}, J.~L., {Slee}, O.~B., {Nelson}, G.~J., \& {Wright},
  A.~E. 1988, \aj, 96, 371, \dodoi{10.1086/114815}

\bibitem[{{Tran} {et~al.}(2016){Tran}, {Cohen}, {Colson}, {Mader}, {Swain},
  {Laity}, {Kong}, {Gelino}, \& {Berriman}}]{Tran2016}
{Tran}, H.~D., {Cohen}, R., {Colson}, A., {et~al.} 2016, in Society of
  Photo-Optical Instrumentation Engineers (SPIE) Conference Series, Vol. 9910,
  Observatory Operations: Strategies, Processes, and Systems VI, 99102E

\bibitem[{{Veyette} \& {Muirhead}(2018)}]{Veyette2018}
{Veyette}, M.~J., \& {Muirhead}, P.~S. 2018, \apj, 863, 166,
  \dodoi{10.3847/1538-4357/aad40e}

\bibitem[{{Veyette} {et~al.}(2017){Veyette}, {Muirhead}, {Mann}, {Brewer},
  {Allard}, \& {Homeier}}]{Veyette2017}
{Veyette}, M.~J., {Muirhead}, P.~S., {Mann}, A.~W., {et~al.} 2017, \apj, 851,
  26, \dodoi{10.3847/1538-4357/aa96aa}

\bibitem[{{Vilhu}(1984)}]{Vilhu1984}
{Vilhu}, O. 1984, \aap, 133, 117

\bibitem[{{Wilson}(1966)}]{Wilson1966}
{Wilson}, O.~C. 1966, \apj, 144, 695, \dodoi{10.1086/148649}

\bibitem[{{Winters} {et~al.}(2018){Winters}, {Irwin}, {Newton}, {Charbonneau},
  {Latham}, {Han}, {Muirhead}, {Berlind}, {Calkins}, \&
  {Esquerdo}}]{Winters2018}
{Winters}, J.~G., {Irwin}, J., {Newton}, E.~R., {et~al.} 2018, \aj, 155, 125,
  \dodoi{10.3847/1538-3881/aaaa65}

\bibitem[{{Winters} {et~al.}(2019){Winters}, {Henry}, {Jao}, {Subasavage},
  {Chatelain}, {Slatten}, {Riedel}, {Silverstein}, \& {Payne}}]{Winters2019}
{Winters}, J.~G., {Henry}, T.~J., {Jao}, W.-C., {et~al.} 2019, \aj, 157, 216,
  \dodoi{10.3847/1538-3881/ab05dc}

\bibitem[{{Wright} {et~al.}(2011){Wright}, {Drake}, {Mamajek}, \&
  {Henry}}]{Wright2011}
{Wright}, N.~J., {Drake}, J.~J., {Mamajek}, E.~E., \& {Henry}, G.~W. 2011,
  \apj, 743, 48, \dodoi{10.1088/0004-637X/743/1/48}

\bibitem[{{Wright} {et~al.}(2018){Wright}, {Newton}, {Williams}, {Drake}, \&
  {Yadav}}]{Wright2018}
{Wright}, N.~J., {Newton}, E.~R., {Williams}, P. K.~G., {Drake}, J.~J., \&
  {Yadav}, R.~K. 2018, \mnras, 479, 2351, \dodoi{10.1093/mnras/sty1670}

\end{thebibliography}
\bibliographystyle{aasjournal}

\end{document}